\documentclass[notoc,amstex]{esty}

\usepackage{epic,epsfig,psfig,axodraw}
\usepackage{subfigure}

\def\del#1{{\bf (deleted text)}}
\def\ghost#1{}

\def\apj#1#2#3{Ap. J. #1, #2 (#3)}
\def\apjl#1#2#3{Ap. J. Lett. #1, #2 (#3)}
\def\physl#1#2#3{Phys. Lett. B#1, #2 (#3)} 
\def\physrev#1#2#3{Phys. Rev. D#1, #2 (#3)}
\def\prl#1#2#3{Phys. Rev. Lett. #1, #2 (#3)}
\def\nucl#1#2#3{Nucl. Phys. B#1, #2 (#3)}
\def\jhep#1#2#3{JHEP #1, #2 (#3)}
\def\nature#1#2#3{Nature, #1, #2 (#3)}
\def\astrop#1#2#3{ Astropart. Phys. #1, #2 (#3)}

\def\ubar#1{\overline{u}_{#1}}
\def\u#1{{u}_{#1}}
\def\v#1{{v}_{#1}}
\def\vbar#1{\overline{v}_{#1}}

\def\dslash#1#2{ \mbox{$#1$ \kern-0.9em \slash \kern0.2em}_{#2} }

\def\dm {Dark Matter }
\def\dmsb {Dark Matter}
\def\cf {\textit{cf. }}
\def\ie {\textit{i.e. }}
\def\eg {\textit{e.g. }}
\def\etc {\textit{etc. }}

\def\mdm {\, m_{dm} \,}
\def\mf {\, m_f \,}
\def\mF {\,m_F \,}
\def\mdmd { \, m_{dm}^2  \,}
\def\mfd { \, m_f^2  \,}
\def\mFd { \, m_F^2  \,}
\def\mdmq { \, m_{dm}^4  \,}

\def\mFq { \, m_F^4  \,}

\def\fl  { \, C_l  \,}
\def\fr  { \, C_r  \,}
\def\fld { \, C_l^2  \,}
\def\frd { \, C_r^2  \,}
\def\flq { \, C_l^4  \,}
\def\frq { \, C_r^4  \,}

\def\U {U}
\def\Ul { U_l}
\def\Ur { U_r}
\def\fzlp { \, f_{\Ul} \, }
\def\fzrp { \, f_{\Ur} \, }
\def\simge{\mathrel{%
   \rlap{\raise 0.511ex \hbox{$>$}}{\lower 0.511ex \hbox{$\sim$}}}}
\def\simle{\mathrel{
   \rlap{\raise 0.511ex \hbox{$<$}}{\lower 0.511ex \hbox{$\sim$}}}}

\def\gtrsim{\mathrel{%
   \rlap{\raise 0.511ex \hbox{$>$}}{\lower 0.511ex \hbox{$\sim$}}}}
\def\lesssim{\mathrel{
   \rlap{\raise 0.511ex \hbox{$<$}}{\lower 0.511ex \hbox{$\sim$}}}}
\def\thickapprox{\mathrel{
   \rlap{\raise 0.511ex \hbox{$\sim$}}{\lower 0.511ex \hbox{$\sim$}}}}

\def\be{\begin{equation}}
\def\ee{\end{equation}}
\def\bea{\begin{eqnarray}}
\def\eea{\end{eqnarray}}
\def\ba{\begin{array}} 
\def\ea{\end{array}}
\def\bc{\begin{center}}
\def\ec{\end{center}}

\newcommand{\UUNIT}[2]{\,{\rm #1}^{#2}}

\begin{document}
\begin{frontmatter}
\title{Scalar Dark Matter candidates}

\author{C. B\oe hm} 
\address{Denys Wilkinson Laboratory, 1 Keble Road, OX1 3RH, Oxford, England, UK}
\author{P. Fayet}
\address{Laboratoire de Physique Th\'eorique de l'ENS\thanksref{labo},
24 rue Lhomond, \\
75231 Paris Cedex 05, France}
\thanks[labo]{\scriptsize UMR 8549, 
Unit\'e mixte du CNRS et de 
l'Ecole Normale Sup\'erieure.}
 
\begin{abstract}
We investigate the possibility that \dm could be
made of scalar candidates and focus, in particular, on the 
unusual mass range between a few MeV's and a few GeV's. 
After showing why the Lee-Weinberg limit 
(which usually forbids a \dm mass below a few GeV's) does not 
necessarily apply in the case of 
scalar particles, we discuss how light candidates ($\mdm < O(\UUNIT{GeV}{})$ )
can satisfy both the gamma ray and relic density constraints. 
We find two possibilities. 
Either \dm is coupled to heavy fermions  
(but if $\mdm \lesssim 100$ MeV, an asymmetry between the \dm particle and antiparticle number 
densities is likely to be required), or \dm is coupled to a new light gauge boson  
$\,U$. The (collisional) damping of light candidates is, in some circumstances, large enough
to be mentioned, but in most cases too small 
to generate a non linear matter power spectrum at the present epoch 
that differs significantly from the Cold Dark Matter spectrum. 
On the other hand, 
heavier scalar \dm particles (\ie with $\mdm \gtrsim O(\UUNIT{GeV}{})$) turn out to be much less constrained.
We finally discuss a theoretical framework for scalar candidates, 
inspired from theories with $N=2$ extended supersymmetry 
and/or extra space dimensions, in which the \dm stability 
results from a new discrete (or continuous) symmetry.
\end{abstract}

\end{frontmatter}


\section{Introduction}

The nature of \dm (DM) is still a challenging question in cosmology. 
While ordinary matter appears to be a very unlikely explanation 
\cite{joe68,macho,davis} to a set of well-known observations, 
the solution based on the existence of a new 
kind of neutral, stable and Weakly Interacting Massive Particles (WIMPs) 
\cite{gunn} 
is still under investigations. The preferred theoretical framework which 
supports such particles, namely supersymmetry,  remains hypothetical 
and possible discrepancies between observations
and some of the Cold Dark Matter 
theoretical predictions are still under debate 
\cite{bmoore,swaters,deblock}.
This makes worthwhile to explore less conventional possibilities,  
as we shall do here.

To satisfy the relic density requirement\,\footnote{
$\Omega_{\hbox{\tiny dm}} \,h^2 \sim \,0.1$, 
where $H= 100 \ h \,\UUNIT{km}{} \UUNIT{s}{-1} 
\UUNIT{Mpc}{-1}$.} as well as other experimental constraints, 
the \dm mass (and, in particular, the mass of the presently favored candidate,
the lightest neutralino), is generally thought to lie between 
$\mdm \sim O(\UUNIT{GeV}{})\,$ (in fact $O(30 \, \UUNIT{GeV}{}$) for neutralinos, assuming unification constraints) 
and $\,O(\UUNIT{TeV}{})$ \cite{delphi,bdd}. 
\,While this range certainly appears natural and promising, 
surprisingly enough, ``light'' annihilating particles 
with a mass below a few GeV's and no significant direct coupling to the 
$Z$ boson do not appear experimentally excluded as yet, as we shall see here. 
This immediately raises the question of their relic abundance.

Annihilating \dm particles with a mass below a few GeV's 
are generally expected to be ruled out because they would overclose 
the Universe \cite{leew}. 
But this argument, based on a relationship between the cosmological 
parameter $\Omega_{dm} h^2$  and the \dm annihilation 
cross section $\sigma_{ann}$, applies only to fermionic candidates 
for which $\sigma_{ann}$ is seen to be proportional to $\mdmd$. 
In the case of scalar particles, however, the situation is different. 
The estimated relic abundance of such candidates may in fact vary considerably, 
depending on if and how these are taken to interact with ordinary matter.
For scalar candidates interacting through Higgs exchanges \cite{zeeetc}
\,-- which interact very weakly with ordinary matter --\,
the resulting weakness of the annihilation cross sections 
tends to lead to an excessively large relic abundance, especially in the case 
of light scalar particles.
We shall discuss here other situations, in which the scalar candidate 
interacts with ordinary matter either through Yukawa 
exchanges of new fermions (such as mirror fermions),
or through the exchanges of a new neutral gauge boson.
The annihilation cross sections can then be largely independent of 
the \dm mass or depend on two low masses (namely the Dark Matter mass and 
the mass of the exchanged particles), so the relationship between 
$\sigma_{ann}$ and $\Omega_{dm} h^2$ eventually turns out to (mostly) 
constrain the mass and coupling constants of the exchanged particles.
We show that in such situations scalar particle masses below a few GeV's
are indeed possible, as far as the relic density argument is concerned.

At this stage it gets necessary to perform a further analysis.
Indeed even if they have the correct relic density light scalar candidates 
could still be ruled out by other astrophysical 
constraints such as, for instance, the precise measurement of 
gamma ray fluxes below a few GeV's. 
By comparing the observed fluxes with those expected from \dm residual 
annihilations into ordinary 
particles,  
it comes out that particles lighter than a few hundred MeV's 
are excluded if their annihilation cross section times velocity 
$(\,\sigma_{\hbox {\tiny ann}}\,v_{rel})$ \, 
behaves mostly as a constant \cite{bens} (unless there is an asymmetry 
between the \dm and anti \dm number densities). 
On the other hand, such particles could escape past 
and possibly present gamma ray detection experiments if 
$\,\sigma_{\hbox {\tiny ann}}\,v_{rel}\,$  was significantly 
dominated, at the freeze-out epoch, by a term proportional to the 
square of the \dm  velocity, \ie if 
$\,\sigma_{\hbox {\tiny ann}}\,v_{rel}\, \thickapprox b \ v_{dm}^2$ 
at the freeze-out epoch (with $b$ a constant).

The aim of this paper is therefore to check whether or not 
scalar Dark Matter candidates can have properties which are 
compatible, simultaneously, with  \vspace{1mm}\\
-- 
the relic density requirement,  \\
--
the measured gamma ray fluxes,  \\
--
and, finally, the present experimental limits from particle physics experiments.

After giving the generic Feynman rules which enter our calculations, 
we discuss, in Section 3, how theories with a new light gauge boson $U$ 
may be constructed, the general properties of $U$ interactions, as well as
possible implications for Higgs bosons.
We compute, in Section 4, the annihilation cross section of light 
scalar candidates into fermion-antifermion pairs.
We then look for the conditions that allow this cross section 
(times relative velocity) 
to give rise to a satisfactory relic density and be significantly 
dominated by a term 
proportional to $v_{dm}^2$ at the freeze-out epoch (which is of interest 
for $\mdm \lesssim 100 \UUNIT{MeV}{}$). 
Because the possible existence of non-chiral Yukawa 
couplings for scalar particles appears very important regarding the relic 
density issue, we also rederive 
for comparison the annihilation cross section  
of fermionic candidates in Section~5. 
Section 6 is dedicated to the question of the collisional damping generated 
by light scalar candidates. To that respect, we compute 
the elastic scattering cross section of \dm with neutrinos and determine 
their associated damping scale.  
We then discuss, in Section 7, 
the case of an asymmetry 
between the number densities of \dm particles and antiparticles, 
as a way to evade the gamma ray constraint on $\mdm \lesssim 100$ MeV.  
Finally, we investigate in Section 8 a theoretical framework 
(inspired from $N=2$ extended supersymmetry and/or theories with 
extra space dimensions) in which a light ``spin-0 gauge boson'' 
(\eg a ``spin-0 photon'') 
\ghost{, that would be stable as a result of a new discrete symmetry,}  
could turn out to be a viable Dark Matter candidate.

\section{\dm couplings \label{dmcoupling}}

\vspace{-5mm}

Although the \dm stability could result from the extreme smallness  
of its couplings to ordinary particles,  we find more elegant to 
impose a new symmetry, \,which may be either discrete or continuous.

The case of self conjugate \dm particles (dm $\equiv$  dm$^*$) 
will be described by introducing a 
new $\,Z_2\,$ discrete symmetry (called $M$-parity and 
denoted $M_p$),  while the case 
dm $\neq$  dm$^*$ will be obtained by imposing other discrete 
or continuous symmetries (\eg $U(1)$), that we call $M$ symmetry 
and denote simply by $M$ (to contrast with $M_p$, dedicated to the 
$\,Z_2\,$ symmetry only). 

In the case of a $M$-parity, 
somewhat similar to $R$-parity in supersymmetric theories 
\cite{pf1}, 
we can distinguish between ``$M$-even'' and ``$M$-odd'' particles (their 
associated fields transforming according to 
$ \phi  \stackrel{M_p}{\longrightarrow} -\ \phi$\,). 
The denomination is slightly different in the case of a 
$M$ symmetry, for which we shall distinguish between 
particles transforming trivially or non-trivially under 
the $M$ symmetry. 
t
If the LMP (\ie the lightest of the $M$-odd particles,  
or the lightest of the particles transforming non trivially under 
the new symmetry) is neutral and uncolored, 
then we end up with a possible \dm candidate. 
This is actually what we shall assume from now on.

\subsection{Direct Yukawa couplings of scalar \dm to matter particles} 

\vspace{-2mm}

Similarly to neutralinos in supersymmetry, we may expect \dm particles 
to be coupled to ordinary 
fermions ($f$) and other particles ($F$). These $F$ particles denote   
new fermions if \dm is bosonic or new bosons if \dm is fermionic. 
The new states $F$ and \dm particles transform in the same way under the new 
symmetry, so we can introduce non-diagonal couplings of the type 
\hbox{dm--$\bar{f}$--$F$} which are expected to 
be invariant under 
\linebreak
\hbox{$M$-parity} or $M$ symmetry, 
whatever the \dm spin (that we shall restrict to be 0 or 1/2 
for simplicity).  We shall set the masses of charged $F$'s to be 
typically above $\sim 100 $ GeV's so as to be compatible with 
unfruitful searches for new charged particles.

Let us now express the corresponding relevant Feynman rules in the case 
of scalar Dark Matter.
The $F$'s denote new spin-1/2 fermions and 
the Yukawa couplings \hbox{dm--$\bar{f}$--$F$} can be written as 
$$\delta \ (\fl \ \,\overline{f_L}\,F_R + \fr\ \,\overline{f_R}\,F_L)
\,+\,h.c.\ \ .$$ This leads to the following Feynman rule:

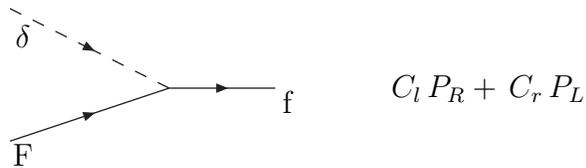
\hspace{4.5cm} 
\begin{picture}(80,80)(0,50)
\ArrowLine(0,70)(60,90)
\DashArrowLine(0,120)(60,90){5} 
\Text(5,65)[]{F}
\ArrowLine(60,90)(100,90)
\Text(105,85)[]{f}
\Text(5,110)[]{$\delta$}
\Text(180,90)[]{$\fl P_R + \fr P_L $}
\end{picture}

where $\,P_L\,$ and $\,P_R\,$ denote the two chiral projectors.

\vspace{0.3cm}

Here $\delta$ stands for the spin-0 \dm field, 
and 
$\,\fl\,$ and $\,\fr\,$ denote the strengths
of its Yukawa couplings to the left-handed 
and right-handed ordinary fermion fields $f_L\,$ and $f_R$, 
\,respectively. 
The left-handed fermion fields $f_L$ transform as members of 
electroweak doublets and their right-handed counterparts as singlets, so 
in the most relevant case for which \dm is an electroweak singlet 
(possibly mixed with the neutral component of an electroweak triplet, 
as we shall discuss later),  
the new fermion fields $\,F_{R}\,$ should transform as members of 
electroweak 
doublets and their left-handed counterparts $\,F_{L}\,$ as singlets. 
In other words, the new fermion fields 
$F$ appear to be {\it \,mirror partners\,} of the ordinary fermion 
fields $f$.

\vspace{3mm}

As also discussed later in Section 8, one can indeed construct models
(possibly but not necessarily inspired from 
extended supersymmetry and/or extra space dimensions), 
in which quarks and leptons (built from left-handed electroweak doublets 
and right-handed singlets), are supplemented by mirror partners
(built from right-handed doublets and left-handed singlets).
Ordinary as well as mirror quarks and leptons acquire their masses, 
as usual, after spontaneous gauge symmetry breaking through
their couplings to one or several electroweak Englert-Brout-Higgs doublets.
Experimental constraints require mirror charged fermions to be sufficiently heavy,
more than about 100 GeV's at least, which may be easily realized 
as soon as their Yukawa couplings to the Higgs 
are taken to be sufficiently large. In addition, mirror neutrinos, 
which did not show up in $Z$ decays, 
should be heavier than about 40 GeV's at least.
This may be realized, for example, by considering them
(at least in a first approximation)
as 4-component Dirac fermions, with large Dirac mass terms
connecting right-handed mirror neutrino fields (members of electroweak doublets),
with associated left-handed extra singlet (``sterile'') neutrino fields
(without introducing large ``Majorana mass terms'' for the latter).

Furthermore, a ``mirror-parity'' symmetry distinguishing 
between the ordinary and mirror sectors prevents ordinary fermions 
to mix with those of the mirror sector. 
Nevertheless, ordinary and mirror fermions may still be coupled together, 
in a gauge-invariant way.
The corresponding Yukawa couplings would then involve
``mirror-odd'' scalars \,-- e.g., in the simplest 
case, a neutral electroweak singlet,  for which one can immediately write 
a gauge-invariant mass term. 
The lightest of these scalars, i.e. in fact the ``LMP'',  would then be stable
thanks to $M$-parity (as soon as it is lighter than all mirror fermions), 
providing, in this way, a possible Dark Matter candidate. 
This particle would interact with ordinary matter, or annihilate in pairs, 
through mirror fermion exchanges with ordinary matter fermions. This 
constitutes, precisely, one of the main subjects of the paper.

\subsection{For comparison, direct Yukawa couplings of spin-1/2  \dm particles.}

Let us now also express, for the purpose of future comparisons, the corresponding 
Feynman rules in the usual case of spin-1/2 \dm particles.
The $F$'s then represent new boson fields.
We make the assumption that they are scalar fields 
(e.g. the squarks and sleptons, in the case of supersymmetry),
for simplicity.
Their non-diagonal Yukawa couplings to \dm and ordinary fermion fields $f$ 
may in principle involve, similarly to the coupling 
$\,\tilde{f}-f-\chi^0\,$ in supersymmetry, the two quantities 
$\,[\,\fl \ \, F \ \overline{f_L} \, \delta_R\,+\,h.c.\,]\,$ and 
$\,[\,\fr \ \, F \ \overline{f_R} \, \delta_L\,+\,h.c.\,]\,$.

\begin{itemize}
\item 
If \dm is a {\it \,Majorana\,} particle (so that $\delta_R = $ 
($\delta_L$)$^c$), and in the simplest case 
for which the $F$ and $\delta_R$ fields 
would appear as eigenstates of the weak isospin $\,T_3$, 
\,the electroweak gauge invariance requires 
that one or the other of the two Yukawa couplings $\fl$ and $\fr$ must vanish identically,
i.e. that the couplings of a Majorana \dm particle with a spin-0 field
$F$ be {\it \,chiral\,}, \ie for example  
$$\fl \ \, F \ \overline{f_L} \, \delta_R \ +\ h.c. \ \ ,$$ which 
gives rise to the rule:

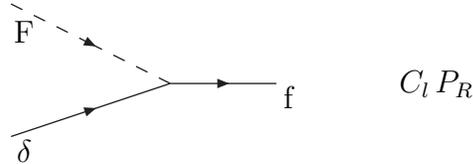
\hspace{4.5cm}
\begin{picture}(80,80)(0,50)
\ArrowLine(0,70)(60,90)
\DashArrowLine(0,120)(60,90){5} 
\Text(5,65)[]{$\delta$}
\ArrowLine(60,90)(100,90)
\Text(105,85)[]{f}
\Text(5,110)[]{F}
\Text(160,90)[]{$\fl P_R $}
\end{picture}

\item 
If \dm is a {\it \,Majorana\,} particle, but 
 $\delta$ (or $F$) appear as mixings of field 
components with
different values of $\,T_3$, then one expects that 
{\it \,non-chiral\,} couplings such as for instance 
$$ \fl \ \, F \ \overline{f_L} \, \delta_R\, + \,\fr \ \, F \ \overline{f_R} \, \delta_L\,  + \,h.c. \ \ ,  $$
be generated, in a way compatible with the (spontaneously broken)
$\,SU(2)\times U(1)\,$ gauge symmetry.
(Such a situation may occur, for example, in supersymmetric theories,
where $F$ would denote a new state resulting from the 
mixing of the two spin-0 sfermion fields $\,\widetilde{f_L}\,$ and 
$\,\widetilde{f_R}$, \,and $\delta$ a Majorana neutralino.) We then get 
the following Feynman rule: 

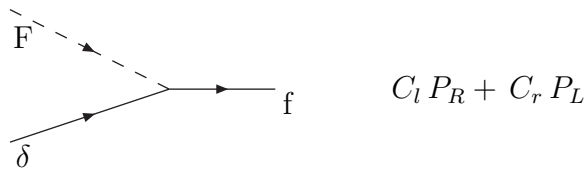
\hspace{3cm}
\begin{picture}(80,80)(0,50)
\ArrowLine(0,70)(60,90)
\DashArrowLine(0,120)(60,90){5} 
\Text(5,65)[]{$\delta$}
\ArrowLine(60,90)(100,90)
\Text(105,85)[]{f}
\Text(5,110)[]{F}
\Text(180,90)[]{$\fl P_R + \fr P_L $}
\end{picture}

\item 
If, on the other hand, \dm is a {\it \,Dirac\,} particle (so that $\delta_L$ and 
$\delta_R$ denote independent field components), 
then the Yukawa couplings of the spin-0 fields $F$ to \dm and ordinary 
fermions may also be written in a non-chiral way as:  
$$ \fl \ \, F \ \overline{f_L} \, \delta_R\, + \,\fr \ \, F \ \overline{f_R} \, 
\delta_L\,  + \,h.c. \ \ .  $$ 
This again gives rise to:

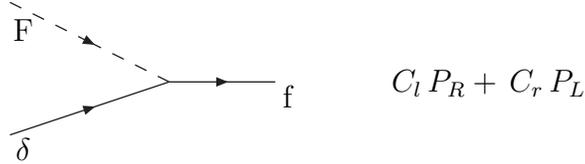
\hspace{3cm}
\begin{picture}(80,80)(0,50)
\ArrowLine(0,70)(60,90)
\DashArrowLine(0,120)(60,90){5} 
\Text(5,65)[]{$\delta$}
\ArrowLine(60,90)(100,90)
\Text(105,85)[]{f}
\Text(5,110)[]{F}
\Text(180,90)[]{$\fl P_R + \fr P_L $}
\end{picture}

\end{itemize}

\subsection{Couplings to $Z$ or Higgs bosons}

\dm particles may interact with ordinary matter through $Z$ boson exchanges, 
as in the case of a neutralino, or scalar neutrino. 
They should then necessarily be heavy, otherwise they would have been
pair-produced in $Z$ decays. A light dark matter candidate 
should have no significant direct coupling to the $Z$ boson,
but it could still interact with ordinary matter through the exchanges 
of other spin-1 gauge bosons (as we shall discuss in the next subsection), 
or of spin-0 Higgs bosons.

In the case of scalar \dm particles interacting with matter 
through spin-0 Higgs boson exchanges, 
as discussed for example in the simple model of 
Ref.\,{\cite{zeeetc}, the weakness of the
Higgs couplings to ordinary fermions leads, in general,  
to a too small annihilation cross section 
and therefore a too large relic abundance of dark matter particles, 
unless one is ready to compensate for that by considering 
a large dark matter/Higgs boson coupling 
(which would tend to make the theory non-perturbative), or to 
assume that \dm is close to in mass to $m_{\rm \tiny Higgs}/2$ 
\,-- and therefore relatively heavy. 
Since Higgs or $Z$ exchanges do not 
really favor light Dark Matter particles, we shall disregard this 
possibility in what follows.

\vspace{5mm}

\subsection{Couplings to a new gauge boson}
\dm particles can also interact 
with ordinary fermions $f$ through couplings to a new neutral 
gauge boson $\U$, should it exist (in fact a $U$ boson has already been proposed 
earlier but in another context \cite{pfu1,pfu2}). 
We shall denote the coupling constant(s) 
between two \dm particles and the $\U$ boson 
\,(i.e. the $\U$-charge of the \dm field)\, 
\,by $C_{\U}$ in the case of a scalar \dm field,  
\,by $C_{\Ur}$ in the case of a Majorana \dm field
(rewritten as a right-handed fermion field) \,and by $C_{\Ul}$ and $C_{\Ur}$  
in the case of a Dirac \dm field, while 
the couplings of the $\U$ boson 
with ordinary chiral fermion fields 
will be denoted by $f_{\Ul}$ and $f_{\Ur}$, respectively. 
We then end up with the following rules:

\hspace{3.3cm}
\begin{picture}(150,80)(100,50)
\DashArrowLine(100,120)(60,100){5}
\DashArrowLine(60,100)(100,80){5} 
\Text(105,115)[]{$\delta^{\star}$}
\Text(75, 115)[]{$p_2$}
\Photon(5,100)(60,100){4}{10}
\Text(105, 85)[]{$\delta$}
\Text(75, 85)[]{$p_1$}
\Text(15,112)[]{U}
\Text(140,100)[]{$C_{U} \, (p_1-p_2)_{\mu}$}
\end{picture}
\hspace{2.3cm}
\begin{picture}(200,80)(140,50)
\ArrowLine(100,120)(60,100)
\ArrowLine(60,100)(100,80) 
\Text(105,115)[]{$\bar{\delta}$}
\Photon(5,100)(60,100){4}{10}
\Text(105, 85)[]{$\delta$}
\Text(15,112)[]{U}
\Text(165,100)[]{$\gamma_{\mu}\, (C_{\Ul} \, P_L + C_{\Ur} \, P_R)$} 
\Text(170,80)[]{or $\,\gamma_{\mu} \,C_{\Ur} \,P_R$}
\end{picture}

\vspace{-5mm}
\hspace{4cm}
\begin{picture}(200,80)(0,50)
\ArrowLine(100,120)(60,100)
\ArrowLine(60,100)(100,80) 
\Text(105,115)[]{$\bar{\rm{f}}$}
\Photon(5,100)(60,100){4}{10}
\Text(105, 85)[]{f}
\Text(15,112)[]{U}
\Text(170,90)[]{$\,\gamma_{\mu} \,(f_{\Ul} \, P_L + f_{\Ur} \, P_R)$}
\end{picture}
\normalsize

\vspace{-5mm}

\section{More on theories 
with a spontaneously broken extra $U(1)$
gauge symmetry}

\subsection{An extra $U(1)$, and its spontaneous breaking}

We shall work within the framework of spontaneously broken 
gauge theories, in which the Standard Model gauge group 
is extended so as to include, 
at least, an extra $\,U(1)\,$ factor. 
The new resulting neutral gauge boson, $U$, acquires a mass from
the spontaneous breaking of the $\,SU(3)\times SU(2)\times U(1)\  \times $ 
extra $U(1)\,$
gauge symmetry into the $SU(3)\times U(1)$ subgroup of QCD $\times$ 
QED\,\footnote{While this extra $U(1)$ was originally introduced 
in connection with spontaneous supersymmetry breaking, 
this initial motivation has faded away
and it should now be considered, independently of supersymmetry.}.
This breaking, and the subsequent mass of the $U$ boson, 
may be obtained explicitly from the non-vanishing vacuum expectation values 
of two electroweak Higgs doublets $\,\varphi_i$, \,such as those present 
in supersymmetric theories\,\cite{pf1}; or of two Higgs doublets + an extra singlet
(that we may call here $\,\sigma\,$), allowing us to increase, if necessary,
the scale $F$ at which the extra $U(1)$ symmetry 
is spontaneously broken, so that the effects of the exchanges, 
or direct production, 
of the new neutral gauge boson $U$ may be sufficiently small\,\cite{pfu1,pfu2}.
Or simply (in the absence of supersymmetry)
by the v.e.v.'s of only one electroweak Higgs doublet ($\varphi$)
and one extra singlet ($\sigma$) \cite{bsp}.
Such additional singlets interacting 
with the Higgs doublets $\,\varphi\,$ 
may be naturally present in many situations, including in particular
non-minimal versions of the supersymmetric standard model \cite{pf1,pfsigma}, 
supersymmetric grand-unified theories, theories with 
extended supersymmetry and/or extra dimensions, ...\,.
In contrast with the Higgs doublet(s) $\,\varphi\,$, 
the extra Higgs singlet(s) $\,\sigma\,$ 
can never couple directly to ordinary quarks and leptons, a point which
will be of importance later.
A general analysis discussing in particular which are 
the possible extra $U(1)$ symmetries which may be gauged\,\footnote{We also
assume that the anomalies associated with the extra $U(1)$ gauge symmetry 
are suitably cancelled, 
which may be easily realized, e.g. by using the $\,B-L$ 
\linebreak and $Y$ symmetry generators 
in the fermionic sector, or by introducing mirror fermions, ...\,.}
(with gauge coupling $g"$), and 
how the associated current $J"$ may mix with the $SU(2)\times U(1)$ current
$\,J_3-\sin^2\theta\,J_{em}\,$ to give the new current $J_U$
and the weak neutral current $J_Z$, and many possible effects 
of a new light gauge boson $U$, 
is given in Ref. \cite{pfu3}.

\subsection{Possible effects on neutral current phenomenology: 
towards a light $U$ boson}

Clearly the question of possible $Z-U$ mixing effects, which might spoil 
the (now very well-known) properties of the $Z$ boson 
and corresponding neutral current $J_Z$, is of primary importance,
and we must make sure that such effects 
be sufficiently small, in the situations we are interested in.
Such a mixing may indeed be absent \,-- as, for example, 
in the simplest situations considered 
in \cite{pf1,pfu1}, when the two Higgs doublets $\,\varphi_i\,$ 
are taken to acquire equal v.e.v.'s (corresponding to $\tan\beta=1$
in modern susy language, or $1/x=1$ in the axion language).

But if we then recover exactly the usual properties 
and neutral current interactions of the $Z$ boson, we still have 
to worry about the effects of  $U$-boson exchanges 
on weak neutral current processes, which have to be
sufficiently small.
In a number of situations, in which the extra $U(1)$ symmetry 
would be broken at or around the electroweak scale $v$, 
for example by two Higgs doublets, 
these $U$-exchange amplitudes would be proportional to $\, g"^2/m_U^{\ 2}\ $ i.e.
precisely to $\,G_F$, 
just as for the standard $Z$-exchange weak amplitudes,
which turned out to be in contradiction with experimental results 
on neutral current processes \cite{pfu1}.

This led us (apart from the obvious possible solution to increase the mass 
of the $U$ so as to make it very heavy)
to pay special attention to situations 
in which the extra $U(1)$ gauge coupling constant $g"$ is taken to be small, 
together with the mass $m_U$ of the new neutral gauge boson $U$;
propagator effects then become essential when estimating $U$ exchange
amplitudes.
This is usually 
(at least for a symmetry breaking scale $\,F \,\simge\,$ electroweak scale)
sufficient to make the $U$-exchange amplitudes 
(proportional to $ \,g"^2/(m_U^2-q^2)$\,)
\,negligible compared to the $Z$-exchange amplitudes
(proportional to $\,(g^2+g'^2)/m_Z^2\,$, i\,.e. to $G_F$), 
\,in high-energy scattering 
experiments (with $|q^2| \gg m_U^{\,2}$ \,\footnote{The  $\,|q^2|\,$ term
then remains essential it the expression of the $U$ boson progator, 
while it may be neglected compared to $m_Z^{\ 2}\,$, 
in the corresponding $Z$ boson propagator.})
\cite{pfu1} \,-- a first success
against too large unwanted effects of the new gauge boson\,!

In the case of experiments performed at lower energies and therefore smaller 
values of the momentum transfer $|q^2| \simle m_U^{\,2}\,$,
on the other hand,
$U$ boson exchanges could still contribute quite significantly or even 
excessively to neutrino scattering cross sections.
Extra scattering amplitudes of neutrinos on electrons, 
fixed by the ratios 
$\ f_{U \nu}f_{U e}/ (m_U^{\,2} - q^2)$ 
\ (denoting by $f_{U \nu}$ and $f_{U e}\,$ the $U$-charges 
of the neutrino and electron fields, respectively),
should be added to the usual standard
model weak-interaction scattering amplitudes, 
which appear with a coefficient $\,2\
G_F\,\sqrt 2\,$.  The neutrino-electron scattering cross section 
has been measured at low energy ($\approx$ 20-50 MeV, 
corresponding to small momentum transfer of a few MeV's), by
the LAMPF and LSND experiments, which found no significant deviation from the
Standard Model \cite{LSND}. The resulting constraint may be written as 
$\ |f_{U \nu}f_{U e}|/ (m_U^{\,2} - q^2) \simle G_F$ or simply
$\ |f_{U \nu}f_{U e}|/ m_U^{\,2} \simle G_F\ $ 
if  the $U$ is heavier than a few MeV's. This
may lead us, subsequently, to consider very small values of the $U$ 
coupling to the neutrinos
(or, possibly, a very small coupling to the electron).


\subsection{A very light $U$ boson does not decouple, 
even in the limit of vanishing gauge coupling\,!}

Once we take this direction of a small $g"$ 
i.e. of a rather weakly coupled and light $U$ boson,
we might na\"{\i}vely think that it is sufficient to consider 
a sufficiently small $g"$ 
so as to get rid of all the unwanted effects of the new gauge boson $U$.
The interactions of this new boson, however,
should not be taken as negligible, 
even in the limit in which {\it the extra $U(1)$
gauge coupling constant $g"$ gets extremely small}\ ! 
\ Indeed an equivalence theorem shows that, in this low mass 
and low coupling limit
(i.e. for small $\,m_U \approx \,g"\,F\,$ and small $g"$, 
\,their fixed ratio defining the scale $F$
at which the extra $U(1)$ symmetry gets spontaneously 
broken\,\footnote{This scale $F$ is
determined by the v.e.v.'s of the various Higgs doublets and singlets ...
responsible for the spontaneous breaking of the extra $U(1)$ symmetry.}), 
the longitudinal polarisation state of the massive spin-1 $\,U$ boson 
does {\it \,not\,} decouple\,! It behaves, instead, 
very much as the massless spin-0 Goldstone boson associated with the 
spontaneous breaking of the extra $U(1)$, now considered as a global 
symmetry \cite{pfu1,pfu2}.
This is quite similar to the equivalence theorem of supersymmetry, 
according to 
which a light spin-3/2 gravitino does {\it \,not\,} decouple in the limit 
$\kappa=
\sqrt{8\,\pi\ G_N}\ \to 0\,$, but interacts 
(proportionally to $\,\kappa/m_{3/2}\,$ i.e. to $1/d\approx1/\Lambda_{ss}^2$)
very much like the massless 
spin-1/2 goldstino of spontaneously broken global supersymmetry\,\cite{pfgrav}.

We should therefore pay a particular attention to these residual interactions 
of the longitudinal polarisation state of the $U$ boson, 
which behaves very much as a (quasi massless) spin-0 Goldstone boson.
In fact it could appear as quasimassless axionlike particle,
which might show up in $\,\psi,\ \Upsilon\,$ 
and (if lighter than $\,\simeq$ 1 MeV) positronium decays
\cite{pfu1,pfu2,pfu3,pfm}!
\,However, if we look at things more closely, the phenomenology of such models 
varies considerably, depending on whether the current 
to which the new gauge boson $U$ couples is purely vectorial
(as e.g. in the one-Higgs-doublet\,+\,one-Higgs-singlet situation detailed in
Ref. \cite{bsp}), 
or has also an axial part, as it occurs in general
when there is more than one Higgs doublet, 
as in supersymmetric theories (cf. the general discussion in Ref. \cite{pfu3}). 
In these latter cases a very light $U$ boson behaves 
very much as a spin-0 axion (in contrast with the simpler case 
of a purely vectorial $U$ current). 
And we then run the risk of having this particle excluded experimentally, 
very much as it happened for a standard axion.

\subsection{The ``invisible $U$-boson'' mechanism}

\vspace{-.3cm}

Fortunately enough a $U$ boson, however, can be made ``invisible'' 
(in particle physics experiments)
if the corresponding extra $U(1)$ symmetry breaking scale $F$
is taken to be sufficiently large 
(as compared to the electroweak scale $\,v\simeq 246$ GeV)
by using, for example, a large Higgs 
singlet vacuum expectation value ($<\!\sigma\!>$) responsible 
for the generation of this large 
symmetry breaking scale ($F \approx \ <\!\sigma\!> \,\gg \,v$) \cite{pfu1,pfu2}. 
(The same mechanism,
incidentally, can also be used to make the interactions 
of the spin-0 axion in particle physics 
almost ``invisible'', 
in correspondence with what happens for the spin-1 $U$ boson,
in agreement with the equivalence theorem mentioned 
earlier\,\footnote{See e.g. footnote {\footnotesize 8} in 
Ref. \cite{pfu1}.}.)
If we denote $\,\sigma\,=\,\frac{h-ia}{\sqrt 2}\,$ this extra Higgs field
transforming non-trivially under the extra $U(1)$ symmetry, 
the longitudinal degree of freedom of the very light spin-1 $U$ boson 
behaves very much as the corresponding --\,also almost ``invisible''\,-- 
(eaten) pseudoscalar Goldstone boson.
The latter, almost identical to $a$ in the large $F$ limit,
is then very weakly coupled to ordinary matter.

\subsection{A purely vectorial $\ U$ current}\label{subsec:vec}

The neutral current $J_U$ to which the $U$ boson couples is obtained 
from the extra $U(1)$ current $J"$, after taking into account 
an additional contribution proportional to 
$\,J_3 -\sin^2\theta\,J_{em}\,$ arising from $Z-U$ mixing effects, 
when such effects are present\,\cite{pfu3,pf5}}.
It involves in general 
an axial part (if there is more than one Higgs doublet) as well as a vector part
which turns out to involve 
(as far as leptons and quarks are concerned) 
a linear combination of the $B$, $L$ and electromagnetic currents.

However there are also simpler situations 
in which the extra $U(1)$ current turns out to be purely vectorial\,\cite{bsp}.
Indeed, to avoid any potential difficulty 
with an axionlike behavior of the $U$ boson
(and having to resort to an extra $U(1)$ symmetry 
broken ``at a high scale'' $F$ significantly larger 
than the electroweak scale), we shall often concentrate 
for simplicity on such situations.
Then no (potentially troublesome) axionlike behavior 
of the longitudinal polarisation state of the 
$U$ boson (associated with non-vanishing axial couplings to quarks and leptons) 
will manifest, and the extra $U(1)$ symmetry is in this case not necessarily 
systematically
constrained to be broken at a scale $F$ higher than the electroweak scale.
Explicit examples have been detailed, 
with the spontaneous gauge symmetry breaking triggered by the  
standard model Higgs doublet $\,\varphi\,$, 
supplemented with an electroweak singlet $\,\sigma\,$ which only transforms 
under the extra $U(1)$ symmetry, 
and does not couple directly to quarks and leptons\,\cite{bsp}.
After mixing effects between neutral gauge bosons are taken 
into account, the $Z$ boson field and $Z$ current 
expressions remain essentially unchanged when one considers, 
as we do here, sufficiently small values 
of the extra $U(1)$ gauge coupling  $\,g"$.
The quark and lepton contribution to the $U$ current 
turns out to be a linear combination of the corresponding 
$\,B$, $L$ and electromagnetic currents \cite{bsp,pfu3}.

\subsection{Implications for the Higgs sector}

Within the above class of models the simplest situation 
is obtained when the standard model Higgs doublet $\,\varphi\,$ 
transforms only under the electroweak $SU(2)\times U(1)$ symmetry 
(and is alone responsible for the $\,W^\pm$ and $Z$ masses), 
and the new singlet Higgs field $\,\sigma\,$ only
under the extra $U(1)$ symmetry 
(and is alone responsible for the $U$ mass), so that no
$Z-U$ mixing effect occurs at this level.
The quark and lepton contribution to the $U$ current appears simply as
a linear combination of the corresponding $B$ and $L$ currents
(such as, for example, $B-L$\,).
The imaginary part ($a$) of the additional Higgs singlet 
$\sigma$ 
generates the extra (``longitudinal'') degree of freedom of the massive 
$U$ boson. Its real part ($h$) corresponds to
an additional neutral scalar (``$CP$-even'') physical Higgs field, which does
not couple directly to quarks and leptons.

The latter singlet field ($h$, given from $\,\sigma \to \frac{h+w}{\sqrt2}\,$) 
may (in general) or may not mix 
with the usual standard model Higgs field 
($H$, \,given from $\,\varphi \to \frac{H+v}{\sqrt2}\,$), 
depending on the existence and magnitude 
of the dimensionless coupling of the $\,|\varphi|^2\,|\sigma|^2\,$
gauge invariant quartic interaction in the scalar potential, 
which may be present in such models (as already in \cite{pfsigma}), 
leading after translation of the Higgs fields to a 
$\,(H+v)^2\,(h+w)^2\,$ interaction term.
In the absence of such a term that would be responsible for 
doublet-singlet mixing effects, 
the usual phenomenology 
of the standard model Higgs boson ($H$)
remains essentially unchanged. The new singlet Higgs boson
($h$), which has a largely arbitrary mass
and no direct coupling to quarks and leptons, is expected to decay instead
into $UU$ pairs through a $\,h\,UU\,$ coupling, 
leading to invisible as well as visible decay modes 
\,-- depending on the invisible (dm dm\, or $\,\nu\bar\nu\,$) or visible
(into $\,e^+e^-$, ...\,, depending on $m_U$) 
decay modes of the $U$ boson.

In general, however, a mixing between the $H$ (doublet) and $\,h\,$ (singlet)
real scalar fields is expected to occur, as indicated above. 
As a result of this $H-h$ mixing, 
the single Higgs boson of the standard model 
$H$ would then get replaced by a 
pair\footnote{Or possibly more, in more elaborate situations.
A similar phenomenon may also occur
in non-minimal supersymmetric extensions of the standard model.}
of Higgs mass eigenstates $\,H_{1,2}\,$.
Each of them would be coupled to quarks and leptons proportionally to their mass
($m_q$ and $m_l$), 
as in the standard model,
times the cosine, or sine, of the mixing angle \,($\xi$)\,
between the two doublet and singlet components 
(originating from the real parts of the $\,\varphi\,$ and $\,\sigma\,$ fields, 
respectively).
The corresponding decay rates of such Higgs bosons $H_i$
into standard model particles 
would then be obtained as e.g. $\,\Gamma_{SM}\,(m_{H_2})\,\cos^2\xi\,$ and 
$\,\Gamma_{SM}\,(m_{H_1})\, \sin^2\xi\,$, in terms of the standard model decay
rates for a Higgs boson of the corresponding mass.
This would of course significantly affect the experimental 
searches for Higgs boson signals. Given the possible hints
for the direct observation of a Higgs signal at LEP\,\cite{LEP}, 
one could even contemplate the possibility 
of having one of these Higgs mass eigenstates 
arond 116 GeV's, for example, and the other at a different mass.

New possible decay modes of these Higgs mass eigenstates $H_i$ into $U$ $U$ pairs,
for example, would also have to be considered in this case.
They would lead to new Higgs decays into e.g. $\,e^+e^-e^+e^-,\ 
\,e^+e^-$ + missing energy, and totally invisible decay modes
from $\,\ U \to \,\nu\bar\nu\,$ or \,dm dm pairs. In addition, the 
quartic $\,hhUU\,$  coupling would also lead to further cascade modes
following the decay of the heavier Higgs boson into the lighter one, 
according to $\,H_2\ \to\,H_1\,UU\,$.

\subsection{Back to Dark Matter candidates}

Being reassured on the possibility 
of reproducing, in various possible ways, 
the successful phenomenology of the Standard Model,
with possibly but not necessarily interesting new features in the Higgs sector,
associated with doublet-singlet mixing and new Higgs decay modes,
we can now proceed and concentrate again on our Dark Matter candidates.

As mentioned earlier the field associated 
with it could be, in the simplest case, a
neutral electroweak singlet but, more generally, it may also appear 
as the result of a mixing 
between the neutral components of singlets, triplets and doublets 
of the electroweak gauge group. However 
a significant contribution from electroweak doublets is usually 
strongly disfavored, as it would generally induce an unwanted diagonal coupling 
to the $Z$ boson, yielding a too large contribution to the invisible decay 
width of the $Z$. Furthermore, charged partners under the electroweak symmetry 
of non-singlet \dm particles  should also be  heavy, i.e. typically 
above $\,\sim$ 100 GeV's.
%

\section{Annihilation cross sections of scalar candidates \label{secannsc}} 

We can now compute the annihilation cross sections of
a pair of \dm scalar particles into a pair  fermion/antifermion 
(the expressions of the squared matrix amplitudes are 
given explicitly in the Appendix). We disregard 
the annihilations into photons which, because 
they involve radiative processes, are generally expected to be 
smaller than the annihilations into fermion pairs. 

We shall see that, in some circumstances, the annihilation cross sections of 
spin-0 particles do not depend significantly on the mass of the incoming 
particles, say in our case the \dm mass. Instead, they depend on other 
parameters, namely the masses and couplings of the exchanged particles. 
This characteristic is 
in fact of crucial importance since it 
allows to evade the lower limit of $\,\sim 2$ GeV's 
given by Lee and Weinberg for spin-1/2 leptons on the 
basis of Fermi interactions (a similar limit, $O(\UUNIT{GeV}{})$, is found also   
in the framework of weak interactions and fermionic candidates).

\subsection{Compatibility with relic density}
\label{comparelic}

The annihilation cross sections of scalar candidates into 
ordinary fermions have already been worked out within a supersymmetric 
framework for sneutrinos \cite{sneutrino1} but we shall rederive their expressions 
because the mass of the outcoming particles (\ie ordinary fermions) 
was neglected, and the terms proportional to $v_{dm}^2$ were disregarded when 
the $v_{dm}$-independent contributions (or, almost equivalently, the S-wave terms) 
were found not to be identically null. 
Note that, to avoid radiative processes that would 
involve both the fermion $F$ and the $U$ boson, we shall consider that 
\dm cannot be coupled to $F$ and $U$ simultaneously. 
This allows us to derive their contributions  
independently. 

\subsection*{$F$ exchanges}

We start by the coupling dm-$f$-$F$, assuming for simplicity 
that the Yukawa couplings 
$\fl$ and $\fr$ are real. We want to check that the cross sections  
associated with \dm candidates lighter than a few GeV's and coupled to fermions $F$ 
can satisfy, for the mass range we are interested in, the simple 
relationship imposed by relic density calculations:  
$\Omega_{dm} h^2 \sim O(\, 10^{-27}-10^{-28}\,)$ \linebreak 
$\UUNIT{cm}{3}\, \UUNIT{s}{-1}/\langle 
\sigma_{ann} \ v_{rel} \rangle $,
or equivalently 
$\langle \sigma_{ann} \ v_{rel}$ $ \rangle \sim O(\, 10^{-26}-10^{-27}$  $\UUNIT{cm}{3}\, \UUNIT{s}{-1})$ with 
$\Omega_{dm} h^2 \sim 0.1$, as determined by several experiments 
\cite{reio1,cospar}.
(Actually the required cross section is expected to be two times smaller 
in the case of self-conjugate \dm particles, as
compared to non-self-conjugate ones.) 
Such values of the cross sections may be obtained, in particular, if  
the annihilation cross section $\sigma_{ann}$ is mostly 
independent of the \dm mass, as we shall see.

When dm $\neq$ dm$^{\star}$, 
the annihilation cross section associated  
with the process dm dm$^{\star} \rightarrow f \bar{f}$ (which 
proceeds via a single ``$t$'' channel, see Appendix \ref{annscni}) 
is given, in the local limit
approximation ($m_F \gg m_{dm}\ge m_{f}$), by:
\begin{equation}  
\label{sann1}                                          
\sigma_{ann} \ v_{rel}\  \simeq  \  \frac{1}{4 \pi} \ \,\sqrt{\,1-\frac{\mfd}{\mdmd}} 
\ \ \bigg(1 - \frac{\mfd}{\mdmd} + v_{dm}^2 \bigg) \ \,\fld \,\frd / m_{F}^2 \ . 
\end{equation}
We quote here only the dominant terms but the full expression can be found in the Appendix.  
For dm$=$dm$^{\star}$ (\ie self-conjugate \dm particles, see  Appendix 
\ref{sec:annscal}), 
one gets essentially the same expression but with 
a larger numerical coefficient:
\begin{equation} 
\label{sann2}
\sigma_{ann} \ v_{rel}\ \simeq \  \frac{1}{\pi}  \ \,\sqrt{\,1-\frac{\mfd}{\mdmd}} 
\ \ \bigg(1 - \frac{\mfd}{\mdmd} + v_{dm}^2 \bigg) \ \,\fld \,\frd/ \mFd \ . 
\end{equation}
The factor four in between (\ref{sann1}) and (\ref{sann2}) 
originates from the fact that one has to take into account  the 
$t$ and $u$ channels plus their interference terms 
(all of them contributing with the same weight).    
Although there is no P-wave term in the case of self-conjugate \dm particles, 
there exists a $v_{dm}^2$ contribution that originates 
from the development of the S-wave term at the second order.

Our results indicate that $\sigma_{ann} \, v_{rel}$ is 
almost independent of the \dm mass under the condition of non chiral couplings.  
As a result, one can obtain $\sigma_{ann} \, v_{rel} \sim O(10^{-26})-O(10^{-27})
\, \UUNIT{cm}{3} \UUNIT{s}{-1}$, 
almost independently of the \dm mass, simply by adjusting the quantity $\fl \fr/m_F$. 
This actually illustrates a situation in which relic density calculations 
appear inefficient to rule out certain values of the \dm mass. 

As an example, $\Omega_{dm} h^2 \sim 0.1$ can be achieved with 
$\,\fl \, \fr  \sim \,0.01-0.1\,$ and  
$\mF \simeq \, O(\,100 \, \UUNIT{GeV}{})- O(\UUNIT{TeV}{})\,$ for any  
value of the \dm mass below a few GeV's. In fact, a more refined  approach 
would consist in using a Boltzmann code \cite{boltz} to determine carefully the 
value of $\langle \sigma_{ann} v_{rel} \rangle$ (and therefore $\mF$ and $\fl \fr$ \etc) 
depending on the freeze-out epoch\footnote{The estimate of the 
freeze-out epoch also depends on the 
annihilation cross section and on the \dm mass but through a logarithm:  
$\mdm/T_{FO} \sim 17 \,+\, \ln (\langle\sigma_{ann} v_{rel}\rangle/10^{-26} \UUNIT{cm}{3} 
 \UUNIT{s}{-1}) \,+\, \ln (\mdm/\UUNIT{GeV}{}) \,+\, \ln \sqrt{\mdm/T_{FO}}$\,. 
\,We then expect the errors on the estimate of 
$\mdm/T_{FO}$ not to affect significantly our conclusions. 
Also it is important to note that the relation 
between the cosmological parameter and the annihilation cross section 
depends linearly  on $\mdm/T_{FO}$, see \cite{griest} for details.}. 
But we think that there is no need to go to this level of accuracy for light
particles, especially since, as we shall see 
(and this is one of the main points of this paper),  
relic density calculations do not provide the most stringent 
constraint on candidates lighter than one hundred MeV's.
 
In the case of {\it \,chiral\,} 
couplings $\fl \fr =0$, the above expressions 
(\ref{sann1}) and (\ref{sann2}) of the annihilation cross sections 
become meaningless\footnote{When 
$\mdm$ is very close but not equal to $m_\mu$,  annihilations 
into muons proceed through a velocity-dependent cross section. 
This actually has no effect on relic density calculations  if 
\dm is coupled to muons as it is coupled to electrons, but 
this could, perhaps, be important for the estimate of gamma ray fluxes 
from \dm particles slightly heavier than the muon 
since photons would be generated mostly through electron-positron pairs.}.
One must return to  
less simplified expressions of the squared matrix amplitudes 
(see Appendix \ref{sec:annscal}, \ref{annscni}) which show 
that the cross sections of candidates lighter than a few GeV's (and with chiral couplings) 
remain significantly too small to allow for an acceptable relic density (unless one increases 
the couplings $\fl$ or $\fr$ up to values which seem very unlikely and 
problematic).

We therefore conclude, at this stage, that 
scalar candidates with $\mdm \lesssim O(\UUNIT{GeV}{})$, 
coupled to heavy fermions $F$, can fulfill the relic density requirement only 
if they have non chiral couplings. 

\vspace{0.3cm}
But such a condition 
could turn out to eventually exclude the possibility of light \dm candidates. 
Indeed, the contribution of a charged $F$ to the muon and electron anomalous magnetic moments, in the case of non 
chiral couplings,  is given by  
$\,\delta a_{\mu, e}^{m_{F} \gg m_{\mu, e}} \simeq 
\frac{\fl \ \fr \ m_{\mu, e}}{16 \pi^2 \, m_{F}}$\,.  
\,Here $\,\frac{\fl \fr}{\mF}\,$ denotes either $\,\left(\frac{\fl \fr}{\mF}\right)_{\mu}$ or  $\left(\frac{\fl \fr}{\mF}\right)_e$, 
depending on whether one estimates the contribution to the muon or electron $g-2$, respectively. 
If not cancelled, $\delta a_{\mu}^{m_{F}}$ has to be lower than a few times $10^{-9}$ 
(extendable up to $5 \ 10^{-9}$) and $\delta a_{e}^{m_{F}}$ lower than a few times $10^{-11}$ (up to 8 $10^{-11}$),  
to be compatible with the differences between the experimental values and the Standard Model predictions. 
Since the quantity $\frac{\fl \ \fr}{m_{F}}$ also enters the 
expressions of the corresponding annihilation cross sections into $\mu^+\mu^-\,$ or $e^+ e^- \,$  
pairs, it is possible to rewrite the cross section as 
\bea
\!\!\!\!\!\!\sigma_{ann} v_{rel} \ \,&\simeq& \ \,\frac{1}{(4) \pi} 
\ \,\sqrt{\,1-\frac{m_{\mu, e}^2}{\mdmd}} \ 
\bigg(\,1 - \frac{m_{\mu, e}^2}{\mdmd} + v_{dm}^2 \,\bigg) \ 
\left(\,\frac{16 \, \pi^2 \ \delta a_{\mu, e}^{\mF \gg m_{\mu, e}} }{m_{\mu, e} } 
\,\right)^2  
\nonumber \\ \nonumber \\ 
&\lesssim& \ \ \ \ \ \ O(\, 10^{-29})\ \UUNIT{cm}{3} \UUNIT{s}{-1}
\nonumber
\eea
(if $\,\delta a_e < 10^{-11}\,$ and $\,\delta a_\mu < 10^{-9}$),
\,which is much lower than the value 
$\sigma_{ann} v_{rel} \sim O(10^{-26}-10^{-27} \UUNIT{cm}{3} \UUNIT{s}{-1})$
required for an acceptable relic abundance.
It is worth noting however that $\sigma_{ann} v_{rel}$ 
can reach $6 \ 10^{-28}$ or 
$6 \ 10^{-29} \ \UUNIT{cm}{3} \UUNIT{s}{-1}$ (for non self-conjugate particles) with the maximum values of  
$\delta a_e$ and $\delta a_{\mu}$ and potentially up to $10^{-26}-10^{-27} \UUNIT{cm}{3} \UUNIT{s}{-1}$ 
if one sums over all the annihilation channels.

Thus, excepted in marginal situations,  
light \dm particles coupled to  fermions $F$ and with a mass $\mdm < O(\UUNIT{GeV}{})$ 
can get the proper relic density only if $a_{\mu, e}^{\mF}$ is cancelled by another contribution. 
This may nevertheless happen if, in addition to the charged $F$, 
there also exists for instance charged scalars ($H^-$) and neutral fermions ($F^0$), 
\cf Appendix \ref{gm2a}. In this case indeed, significantly larger values of the annihilation cross section
would be allowed, but at the price of a fine tuning.
In fact such a cancellation may also occur by introducing a new (pseudo)scalar particle. 
For instance, in the framework of theories originating from $N=2$ extended
supersymmetry or extra dimensions (\cf Section 8),
the \dm candidate is supplemented by another neutral spin-0 particle,
which may also be relatively light. 
Both the \dm and the other spin-0 particle would have non-chiral couplings to fermion/mirror fermion pairs 
but one would be a scalar while the other would be a pseudoscalar. 
Each separate contribution to the muon or electron $g-2$ 
is expected to be too large, but their sum would be, on the other hand,  
naturally small. Large values of the
annihilation cross section would then be allowed, without being in conflict with $g-2$
constraints.
\ghost{(One might also imagine, for example, that \dm particles could be more strongly 
coupled to quarks than to leptons, so that their annihilations 
would proceed mostly towards hadrons (pions) if they are heavier 
than about 200 MeV's,
the $\,g-2\,$ constraints 
then becoming irrelevant.)}
In all these cases, the particle spectrum 
required to get an acceptable relic abundance 
should be within the reach of future LHC experiments.

\subsection*{$U$ boson}

Let us now estimate the efficiency of the \dm annihilation process 
through the $s$-channel production of 
a gauge boson $\U$. We assume that the direct 
\dm coupling to the $Z$ is null (or extremely small).  
As mentioned previously, 
the coupling between the $U$ boson and ordinary fermions 
($\U-f-\bar{f}$) can be written as $\gamma^{\mu} (f_{\Ul} P_L + f_{\Ur} P_R)$ 
and the coupling between the $U$  and 
the \dm particles as $\,C_{\U}$.
The corresponding annihilation cross section is then given by 
\be
\label{sannz}
\!\!\!\!\!\!\!\!\!\!
\sigma_{ann} \ v_{rel} \ \simeq \
v_{dm}^2 \ 
\hbox{$
\sqrt{1-\frac{m_f^2}{m_{dm}^2}}
$} \ \
\hbox{\large$
\frac{C_{\U}^2 \ \left[4 \mdmd (f_{\Ul}^2 + f_{\Ur}^2) - 
\mfd (f_{\Ul}^2 - 6 f_{\Ul} f_{\Ur} + f_{\Ur}^2) \right]}
{12 \,\pi \ (m_{\U}^2 - 4 \, \mdmd)^2}\ \ 
$}
\ee
(we assume $m_U > 2\,\mdm$). 
The interesting feature here is that the S-wave term is naturally
suppressed (as noted in \cite{sneutrino1} for sneutrinos), 
due to the derivative nature of the 
couplings between the $\U$ boson and the \dm particles.

If the $\U$ boson  is very heavy (more than a few 
hundred GeV's) and the \dm mass lower than a few GeV's, 
then one expects the cross section given in eq.\,(\ref{sannz}) to be too small to allow 
for an acceptable relic abundance. On the contrary, if the $\U$ boson turns out 
to be light (less than a few GeV's), one may obtain the appropriate value of the 
annihilation cross section provided the couplings $C_U$ and 
$f_{U_l}, \ f_{U_r}$  
be sufficiently small. This is actually fortunate since the smallness 
of the coupling $U-f-\bar{f}$ is also needed to fulfill, in particular,  
experimental limits from the muon and electron's anomalous magnetic moments 
(see Appendix \ref{gm2a}). 
Basically, $f_{U_l}$ and $f_{U_r}$ associated with either $U-\mu-\bar{\mu}\,$ 
or $U-e-\bar{e}\,$  
should be smaller than about 
$3 \, 10^{-6} \, \left(\frac{m_U}{\UUNIT{MeV}{}}\right) \, (\frac{\delta a_{\mu}}{10^{-9}})^{1/2}$ 
(in fact $3 \, 10^{-4} \, \left(\frac{\delta a_{\mu}}{10^{-9}}\right)^{1/2}$ if $m_U < m_{\mu}$)  
or $7 \, 10^{-5} \, \left(\frac{m_U}{\UUNIT{MeV}{}}\right) \, \left(\frac{\delta a_e}{10^{-11}}\right)^{1/2}$ 
(assuming $f_{U_l}=f_{U_r}$ and depending whether $m_U$ is greater than $m_{\mu}$ or not), 
in order to be compatible with both the muon and electron's $g-2$.

Taken at face value, the couplings $f_{U_l}$  (and $f_{U_r}$) 
can give rise to an annihilation cross section 
(into a pair  muon-antimuon or electron-positron depending on the \dm mass) 
of the order of $10^{-26}-10^{-27} \UUNIT{cm}{3} \UUNIT{s}{-1}$ 
provided that the $U$ boson happens to be more strongly coupled 
to the \dm 
than to ordinary fermions. As an illustration, to obtain
$\,\Omega_{dm} h^2 \sim 0.1\,$ with, for instance, $f_{U_l} \sim f_{U_r} \sim 4 \ 10^{-4}$ 
($\delta a_e \simeq 3 \ 10^{-12}$), 
$\mdm \sim 4$ MeV and $m_U \sim 10$ MeV, $\,C_{\U}$ should be of the order of 
$\sim (2.5-8) \ 10^{-3}$. More generally, one obtains the correct relic density 
if $\ C_U \ f_{U_l} \sim \,(3-12) \ 10^{-8} \ \left(\frac{m_U}{\UUNIT{MeV}{}}\right)^2 \ \left(\frac{\mdm}{\UUNIT{MeV}{}}\right)^{-1}$.

In fact, two expressions of $C_{U}$ are displayed in Table \ref{table}, 
depending on 
\linebreak
whether 
dm dm $\rightarrow e^+ e^-$ or dm dm $\rightarrow \mu^+ \mu^-$ is the dominant 
channel.  If \linebreak
$\langle \sigma_{ann} \ v_{rel} \rangle_{dm \,dm \ \rightarrow \, \mu^+ \mu^-} = 
\langle \sigma_{ann} \ v_{rel} \rangle_{dm \,dm \ \rightarrow \, e^+ e^-}$, 
then the product 
$(f_{U_l} \, C_U)_{\mu}$  should be equal to $ (f_{U_l} \, C_U)_e$. Since $C_U$ is related to \dm only, 
one gets $(f_{U_l} )_{\mu} = (f_{U_l})_e$ (which can be further translated into  
a relationship between $\delta a_e$ and $\delta a_{\mu}$).

In conclusion, scalar \dm particles can be significantly 
lighter than a few GeV's (thus evading the generalisation of the Lee-Weinberg 
limit for weakly-interacting neutral fermions) 
if they are coupled to a new (light) gauge boson or to new 
heavy fermions $F$ (through non chiral couplings and potentially 
provided one introduces additional particles). 
Let us now determine if they can be compatible with the observed production of gamma rays.

\subsection{Compatibility with gamma rays}

In \cite{bens}, it was shown that 
if light \dm candidates ($m_{dm}<O(100)\UUNIT{MeV}{}$) have a roughly
``constant'' annihilation cross section of 
$\langle\sigma_{ann} \ 
v_{rel} \rangle \sim O(10^{-26}- 10^{-27}\UUNIT{cm}{3} \UUNIT{s}{-1})$ 
(as needed to satisfy the relic density 
requirement), they would yield too much gamma rays as
compared with  present 
observations\,\footnote{We assume here that there is no asymmetry 
between the \dm particle and antiparticle number densities so that 
one can use residual annihilations in galaxy clusters or centres of 
galaxies as a constraint on the properties of light \dm particles.}. 
They would be allowed by COMPTEL, OSSE and EGRET data \cite{egret}, 
on the other hand,
if their cross section (times the relative velocity) at the moment of the residual annihilations  
was $\sim \, 10^{-31}$ ($10^{-29}$ or 
$10^{-28}) \ \UUNIT{cm}{3} \UUNIT{s}{-1}$ for 
$\mdm \sim 1 \ (10 \ \mbox{or} \ 100)$ MeV respectively.  
This actually suggests that light \dm particles with a constant cross section 
are ruled out while particles that would mainly annihilate 
through a velocity-dependent cross section would be allowed. 
Thereby, this requires that the \dm velocity-independent cross section 
be at least $\,\sim 10^{5}$ (resp. $10^{3}$ or $10^{2}$) times smaller than the velocity-dependent cross section  
estimated at the freeze-out epoch ($t_{\small{FO}}$),  
for $\mdm \sim 1 \ ( \ 10 \ \mbox{or} \ 100)$ MeV (and 
$\langle \sigma_{ann} \ v_{rel} \rangle \sim 10^{-26} \UUNIT{cm}{3} \UUNIT{s}{-1}$). 
Let us now check if the cross sections displayed above have 
the proper characteristics.

\subsection*{Residual annihilations through $F$ exchanges} 

\vspace{-5mm}
We have seen previously that the coupling dm-$\bar{f}$-$F$ should be non chiral in 
order to get the correct relic density. However, in this case, the 
S-wave term in the cross section 
appears of the same order of magnitude than 
the velocity-dependent term. Therefore light scalar 
candidates (with $\mdm \lesssim O(100)$ MeV) that would get the proper relic density, 
are likely to be excluded because they would yield a too large gamma ray production 
(unless one invokes an asymmetry between the \dm particle and 
antiparticle number densities since, in this case, no significant residual annihilations 
are expected). Only searches for new particles in accelerator experiments would then 
constrain the properties of charged $F$'s (and of the additional 
spectrum needed to cancel its associated $g-2$ contributions).

\vspace{-3mm}
\subsection*{Residual annihilations through $U$ production}

\vspace{-5mm}
The gamma ray constraint appears much easier to satisfy with a $U$ boson 
because the cross section is 
naturally ``S-wave'' suppressed (the constant term  vanishes identically). 
The value of the annihilation cross section inside the galactic centre  
is expected to be suppressed by the square of the \dm velocity. 
We therefore expect the cross section to be 
roughly $3 \, 10^{-6}$ times what is needed at the freeze-out epoch, 
leading to a radiative flux 
$\ \phi \sim  2 \ 10^{-5} $ $\left(\frac{\mdm}{\UUNIT{MeV}{}}\right)^{-2}
 \UUNIT{cm}{-2} \UUNIT{s}{-1}\ $ 
for $\mdm \in O(1-100)$ MeV (using a NFW profile \cite{NFW}). 
This flux would be smaller with another \dm halo profile. As a result, $O(1-100)$ MeV 
particles could indeed have escaped previous gamma ray experiments, although masses  
$\mdm \sim O(\UUNIT{MeV}{})$  seem (under the condition of a NFW profile) very close to 
experimental limits.

To summarize, the production of 
gamma rays provides a much stronger constraint on scalar particles 
with $\mdm \lesssim 100 \UUNIT{MeV}{}$  
and on the nature of their interactions,
than relic density calculations. This study indeed indicates that 
i) annihilations through $F$ exchanges are forbidden unless there exists 
an asymmetry between the \dm and anti \dm number densities and ii) 
a gamma ray signature from the galactic centre at low energy could be due to the 
existence of a light new gauge boson. The range beyond a few hundred 
MeV's, on the other hand, is less constrained by gamma ray fluxes (whether 
\dm is coupled to a $U$ boson or a $F$ particle) 
but some constraints on the production of $D + ^3 H_e$ \cite{mcdonald} may also require a 
velocity-dependent cross section.

Let us  mention a point concerning the existence of 
the $U$ boson that might be important especially since WMAP's result  
\cite{mapreio}. Because the annihilation 
cross section appears to be S-wave suppressed, the residual annihilations of 
light \dm are expected to be important when the \dm velocity 
and number density are the highest. We would therefore expect a ``gamma ray'' 
signature in the centres of galaxies or clusters of galaxies at low 
redshift  but there might also exist a signature at 
larger redshift where the (mean) \dm number density is supposed to be 
bigger (due to redshift effects). This signature, should it exist, could be a kind of early reionization, or 
perhaps an indirect source of reionization. Its importance 
would depend on the \dm velocity  and number density that is embedded 
in structures formed at a given redshift. Although we did not perform 
any calculations, it is worth mentioning this effect 
because it might allow to constrain the scenario of 
light \dm particles coupled to a $U$ boson.

\section{Annihilation cross sections of spin $\frac{1}{2}$ candidates}

Although this is not the purpose of our paper, we now return for comparison 
to a more usual situation where \dm is made of spin-$\frac{1}{2}$ particles. 
Due to a Lee-Weinberg-type limit for ``weakly-interacting'' 
massive leptons, we expect fermionic candidates with a mass smaller than a few GeV's 
to be ruled out. If coupled to a scalar $F$, then the 
annihilation cross section of spin-1/2 \dm particles is expected to be 
proportional to $\mfd/\mFq$ or $\mdmd/\mFq$. This is actually 
much too small to prevent them from overclosing the Universe 
(unless one considers the exchange of a relatively light neutral 
scalar $F$ that would not be coupled to the $Z$ boson). 

Calculations displayed in the literature 
have shown that for Dirac fermions
both the S and P-wave cross sections are proportional
\footnote{Keeping the dominant terms, the annihilation cross section of spin-1/2 Dirac fermions 
is given by: 
$$
\sigma_{ann} \ v_{rel}  \ \sim \ 
\frac{1}{32 \ \pi\ {m_{F}}^4} \ \ \sqrt{\,1-\frac{\mfd}{\mdmd}}
\ (\fld + \frd)^2 \, \mdmd \, \bigg\{ \, 1 + \, \frac{7 v_{dm}^2 \, }{3}\, \bigg\}
\ .
\nonumber 
$$
}
to $\mdmd/\mFq$ (assuming that the exchanged 
particle, namely a scalar $F$, is much heavier than the \dm and quoting only the most relevant 
terms)   
while for Majorana fermions, the S-wave term is proportional to $\mfd/\mFq$  
and the P-wave term to $\mdmd/\mFq$, unless   
the couplings are non-chiral (i.e $\fl \fr \!\!\neq 0$) \cite{xoann}.  
In this case, indeed, it is known that 
the S-wave annihilation cross section of Majorana particles is 
proportional to $\mdmd/\mFq$ instead\,\footnote{For instance, in 
supersymmetry, non chiral couplings as the result of a mixing angle 
between the two stop fields $\widetilde{t_L}$ and $\widetilde{t_R} \,$, 
have to be taken into account in the estimate of the annihilation cross 
section 
of neutralinos into $t \bar{t}$ (through a stop exchange).} of $\mfd/\mFq$.

Since spin-1/2 Majorana \dm particles  have 
the following annihilation cross section 
(keeping only the dominant terms): 
\bea
\!\!\!\!\!\!\!\!
\sigma_{ann} \ v_{rel} \  &\sim &\ 
 \frac{1}{32 \ \pi\ {m_{F}}^4} \ \ \sqrt{\,1-\frac{\mfd}{\mdmd}} \ \
\bigg\{\ \bigg(\,2 \fl \fr \mdm + (\fld + \frd) \mf \,\bigg)^2 \ 
\nonumber \\ \nonumber \\
\ && \ \ \ \ \ \ \ + \ \ \frac{\ 4 \ v_{dm}^2}{3} \ \ \bigg(
\,\mdmd \ (2 \flq \ + \ 9 \fld \frd \ + \ 2 \frq) \,\bigg) \,+\, ...\ \bigg\} \ ,
\nonumber\\ 
\label{major} \nonumber
\eea
one should keep the S-wave term in a situation where the 
couplings are non chiral. This is expected to be important 
for \dm particles  much heavier than either the tau 
(or the bottom quark, but slightly lighter than 
the top, \ie $m_{\tau, \ b} \ll \mdm \lesssim m_t$), 
or heavier than the top ($\mdm > m_t$).   
In these cases, indeed, the gamma ray flux -- 
proportional to the \dm residual annihilation 
cross section -- is expected to be larger than previous estimates 
based on a cross section 
proportional to $\mdmd v_{dm}^2/\mFq$ or $\mfd/\mFq$.

More generally, the S and P-wave terms that we find by assuming non chiral couplings 
appear, at the freeze-out epoch, somewhat larger than the P-wave term 
estimated in the presence of chiral couplings of similar magnitude. This 
suggests that the lower bound on the \dm mass obtained 
in the case of non chiral couplings is somewhat smaller than 
the bound derived when the couplings are chiral. However, 
despite this characteristic, the lower limit on the \dm mass is still  expected to 
be of a few GeV's (assuming $\mF > m_W$, which could perhaps be contested 
if the primordial annihilations mainly proceed into neutrinos through neutral $F$ exchanges). Therefore, the mass range 
below a few GeV's is very likely to require scalar \dm particles (rather than spin-1/2 particles),  
assuming \dm has indeed to annihilate to get the proper relic density.

\section{Collisional damping effect}

\dm is generally thought to interact weakly enough to be 
considered as collisionless regarding  structure formation 
\cite{gunn}. However, 
a more detailed study shows that even 
\dm particles with weak interactions yield 
damping effects in the linear matter power spectrum \cite{bfs,brhs} 
at a given scale.  The issue is then to determine whether this scale 
is of cosmological interest or not. 

The answer to that question is related to the quantity 
$\,\sigma_{el}/(\mdm/\UUNIT{GeV}{})$, 
denoting the ratio of the \dm elastic scattering cross section to its mass (measured in GeV). 
If this ratio is equal to a 
certain value $\hat{\sigma}$ (that we shall precise later), 
then the \textit{linear matter power spectrum} P(k) differs from the ``Cold Dark Matter'' spectrum 
at all scales below a characteristic length (namely the damping scale) $l$, essentially determined by 
$\hat{\sigma}$. This means that there is a relationship between the damping scale  $l$ and 
$\hat{\sigma}$, bearing in mind that $l$ has to be lower than the scale $l_{struct}$ of the smallest 
primordial structures  that have been observed. 
What are the values of $\hat{\sigma}$ which induce cosmological effects 
in the linear matter power spectrum? Are those values possible 
in the case of light \dm particles? This is what we shall now determine. 

Since the elastic scattering cross sections of light \dm candidates 
cannot be treated independently of the annihilation cross sections,   
we shall consider only values of the  coupling constants,  
$m_{F}$ or $m_{\U}$ that are compatible with relic density calculations. 
 
\subsection{Constraints on \dm interactions} 
In the picture commonly adopted nowadays, 
\dm would belong to either the Cold or Warm \dm scenarios. 
The latter is however less popular  because of the lack of candidates and, presumably,  
because it relies on a free parameter, the \dm mass.  
A lower limit on $\mdm$  
can be obtained from the comparison of the free-streaming length with 
the scale $l_{struct}$. 
(Basically, $m_{dm} \geq 0.75 \UUNIT{keV}{}$ preserves  
all scales beyond $l_{struct} \geq 0.16 \UUNIT{Mpc}{}$ 
\cite{spergel}, 
but the WMAP's results now suggest that the values  
$m_{dm} \lesssim 0.75 \UUNIT{keV}{}$ and even perhaps $\lesssim 10 \UUNIT{keV}{}$ are 
ruled out \cite{reio1,reio2}, although this can be discussed too \cite{julien}).

In a realistic description, where one takes into account the 
\dm interactions, the parameter $\mdm$ is 
associated with another 
parameter, namely the \dm interaction rate. As a consequence, 
the free-streaming scale for the kind of particles we are interested in 
\begin{itemize} 
\item depends on both $\mdm$ and $\sigma_{el}$ (the \dm elastic scattering cross section that 
fixes its  thermal decoupling);  
\item is supplemented by 
an additional damping length, namely the collisional damping scale, 
more directly related to \dm interactions.
\end{itemize} 
This additional scale can be split into two parts, namely a self-damping and 
induced-damping 
contributions, which respectively describe the influence of  
\dm properties, 
and that of other species, on primordial fluctuations. 

Because the induced damping actually corresponds 
to the damping acquired by 
a species $i$ (\eg electrons, photons, or neutrinos) which is further 
transmitted to the \dm fluctuations if the coupling dm$-i$ is strong enough, 
there exists a relationship between the elastic cross 
section  $\sigma_{el(dm-i)}$, the \dm mass and the scale $l$ at which the damping is 
expected to be of noticeable importance \cite{bfs}.  

For the two species $i=\gamma,\nu$ which are 
expected to communicate the largest damping effects (and therefore 
the most stringent constraint on the cross section) \cite{bfs,brhs}, 
this relationship can be displayed as:
\begin{equation}
\sigma_{el(dm-\nu)}\ \lesssim \ 3 \  10^{-35}\, \UUNIT{cm}{2} \ 
\left(\frac{m_{dm}}{\UUNIT{GeV}{}} \right) \ 
\left(\frac{l}{100 \,\UUNIT{kpc}{}}\right)^2 \label{dmnu}
\end{equation} 
and 
\begin{equation}
\label{dmga}
\sigma_{el(dm-\gamma)} \ \lesssim \ 6 \  10^{-34} \,\UUNIT{cm}{2} \ 
\left(\frac{m_{dm}}{\UUNIT{GeV}{}}\right) \ 
\left(\frac{l}{100 \,\UUNIT{kpc}{}}\right)^x
\end{equation} 
with $x \sim 1$, the damping scale being maximal when 
$l = l_{struct} \simeq 100$ kpc. 
We then obtain $\,\hat{\sigma} \sim 3 \ 10^{-35} \UUNIT{cm}{2}\ 
\left(\frac{l}{100 \, \UUNIT{kpc}{}} \right)^2 $ when \dm is coupled to neutrinos 
and 
$\,\hat{\sigma} \sim 6 \ 10^{-34} \UUNIT{cm}{2} \ 
\left(\frac{l}{100 \, \UUNIT{kpc}{}}\right) \,$ when \dm is coupled to photons.

The cross section (\ref{dmnu}) appears, at a first glance, 
very close to that obtained from the relic density condition, for particles 
with $\mdm \sim 10 \UUNIT{MeV}{}$ and  $l = 100 \UUNIT{kpc}{}$
(suggesting a surprising large damping effect) 
but one has to check that the elastic scattering cross section 
$\sigma_{el(dm-\nu)}$ indeed behaves in the same way as the annihilation 
cross section $\sigma_{ann}$.

The constraint on dm-$\nu$ interactions comes from a new 
damping effect which can be seen as a mixing 
between the collisional damping and free-streaming effects \cite{bfs}. 
It is such that any freely-propagating species $i$ ($i=\nu$ in our case) 
can communicate its free-streaming damping to the \dm fluctuations provided that 
the species $i$ happens to have interactions with \dm which  
change the velocity of the \dm fluid  but not\,\footnote{Such a mechanism is 
irrelevant when dealing with the Silk damping effect because the baryons and 
photons thermally decouple (\ie become collisionless) at the same time due 
to recombination.} that of the fluid $i$ (say differently, such a 
coupling must not change the thermal decoupling of species $i$).
The mixed damping is therefore at work only if the DM-neutrino 
interactions can decouple at a temperature below $\sim$ 1 MeV.  

Equations (\ref{dmnu}) and (\ref{dmga})    
are given at specific times, namely the thermal decoupling of \dm 
from neutrinos and photons (say $t_{dec(dm-\nu)}$ and $t_{dec(dm-\gamma)}$  
respectively). If 
$\sigma_{el(dm-\nu)}$ and $\sigma_{el(dm-\gamma)}$,  estimated 
in a given particle physics model, appear to be 
temperature dependent (\ie $\ \sigma_{el(dm-i)} \propto \ T^n$), 
then a more relevant constraint turns out to be given by 
the parameter $b_{el}$, defined by $\ \sigma_{el(dm-i)} \ c \ = \ b_{el(dm-i)} \ T_{dec(dm-i)}^n \ $,  
where $n$ is fixed by the model.

From now on, we shall focus on the neutrino induced-damping effect. 
We disregard the \dmsb-photon elastic scattering cross section (based on box and triangle diagrams) 
in order to simplify our calculations. This one is expected to vanish with 
the photon energy $E_{\gamma}$, for $E_{\gamma}$ lower than the \dm mass,
due to gauge invariance and low energy theorems.
We shall see that, within reasonable assumptions, the elastic scattering cross 
section of scalar \dm particles with 
neutrinos at the \dm thermal decoupling also appears to vanish with 
the neutrino energy (in fact it is proportional to the square of the neutrino energy, 
$\ \sigma_{el(dm-\nu)} \propto \ T^2$, assuming \dm is still in 
equilibrium with this species). The estimate of 
the neutrino induced damping effect is then expected to provide a damping scale either 
comparable or potentially larger than the photon constraint (although one should do in 
principle the comparison).
%

To estimate the neutrino induced-damping scale, we need first to express $b_{el(dm-\nu)}$ explicitly. 
By using eq.\,(\ref{dmnu}) and  the relationships between 
i) the cross section and the comoving interaction rate: 
$$ \tilde{\Gamma}_{dec(dm-i)} \ \simeq \ \sigma_{el(dm-i)} \ c \ \tilde{n}_{dm} \ \frac{\rho_{i}}{\rho_{dm}} 
\ \simeq \ \sigma_{el(dm-i)} \ c \ \tilde{n}_{i} \ \frac{T}{\mdm}\ , 
\ \ \ \ \mbox{if}\  \mdm > T$$ 
(with $\tilde{n}_{i} = \ n_i \ a^{3}$), 
\ ii) the comoving 
interaction rate and the scale factor in the radiation dominated era, 
$$ \tilde{\Gamma}_{dec(dm-i)} \ \thickapprox \ 2 \ 10^{-20} \ a_{dec(dm-i)} \ 
\UUNIT{s}{-1},$$ and finally \ iii) the scale factor and the temperature: 
$$a_{dec(dm-i)} \ \simeq \ \frac{T_0}{T_{dec(dm-i)}}\ \ ,$$
one gets the relation: 
$$
b_{el} \ \thickapprox \  2.5 \ 10^{28} \, \left(\frac{\mdm}{\UUNIT{GeV}{}}\right)^{-1} \, \left(\frac{\sigma_{el(dm-i)}}{\UUNIT{cm}{2}} \right)^2 
\UUNIT{cm}{5} \, \UUNIT{s}{-1}\ \ ,
$$
which turns into the following constraint: 
\begin{equation} 
\label{tdeux}
b_{el(dm-\nu)} \ \lesssim\  2 \  10^{-41} \ 
\left(\frac{l_{struct}}{100 \,\UUNIT{kpc}{}}\right)^4  \, 
\left(\frac{\mdm}{\UUNIT{GeV}{}}\right) \, \UUNIT{cm}{5} \, \UUNIT{s}{-1}\ .
\end{equation}

We now have to compute $b_{el(dm-i)}$ in the case of light scalar \dm candidates to determine 
whether they can induce changes in the  linear 
matter power spectrum on visible scales.

\subsection{Elastic scattering cross section of scalar \dm with neutrinos} 

Once again, we shall assume that \dm is not coupled to a fermion  $F$ and a $\U$ boson 
simultaneously. 
The elastic scattering cross section of non self-conjugate 
\dm (dm$\neq$dm$^{\star}$) with a (left handed) neutrino 
turns out to be given by: 
\be
\sigma_{\tiny{el(\mbox{dm}-\nu)}} \ \thickapprox \ \,
\frac{\fl^4}{8 \ \pi \ \mFq} \ \,T^2  
\ee
(within the local limit approximation for which $\mdm \ll m_F$ and assuming 
that the energy of the scattered neutrino is the same as that of the incoming one). 
In constrast, due to the relative sign between the $t$ and $u$-channel 
(see Appendix \ref{nusc}), 
the cross section $\sigma_{el(\tiny{\mbox{dm}-\nu)}}$ of 
self-conjugate scalar \dm particles
vanishes identically. Consequently, self-conjugate scalar candidates 
do not suffer from neutrino-induced damping effects at all.
%

A rough estimate indicates that 
$\,b_{el} \thickapprox \,5 \ 10^{-55} \flq (\mF/100 \UUNIT{GeV}{})^{-4} \UUNIT{cm}{5} \UUNIT{s}{-1}$ 
for $\mbox{dm}\neq\mbox{dm}^{\star}$. If  $\mF \sim 100 \UUNIT{GeV}{}$,  
$\fl \approx 1$ and $1 \lesssim \mdm \lesssim 100 \UUNIT{MeV}{}$, 
all the scales 
smaller than $\lesssim 10-0.3 M_{\odot}$ are damped respectively. 
Such a damping scale is actually too small to be constrained by present observations or tested by  arguments on the 
reionization epoch \cite{mapreio,julien,reio2} so it might be very difficult to exclude 
such particles from their linear and non-linear  matter power spectra. This collisional damping scale would be 
in fact even smaller if  $\fl \ll 1$ or $\mF \gg 100$ GeV. (One could consider,  
however, smaller values of $m_F$ provided one checks 
that they indeed evade limits notably on neutral supersymmetric particles.)  
 
For comparison, the free-streaming scale 
(or, almost equivalently in this case, the self-damping scale), for particles 
which are likely to annihilate, behaves like 
$ l_{fs} \thickapprox 190 \UUNIT{kpc}{} \left( \frac{\mdm}{1 
\UUNIT{MeV}{}} \right)^{-1/2} \left( \frac{a_{dec(dm)}}{10^{-4}}\right)^{1/2}$
(if they thermally decouple in the radiation dominated era 
\ie if $\,a_{dec(dm)} < 10^{-4}$). 
This expression is different from the well-known free-streaming formula 
because we are dealing with particles which have non-negligible 
interactions, \ie for which the thermal decoupling epoch occurs 
after the non relativistic transition. 

Before using this formula, one can check that light particles (with $\mdm \gtrsim O(\UUNIT{MeV}{})$) indeed thermally 
decouple after their non relativistic transition 
\linebreak
(so that they can potentially annihilate). 
The elastic cross section of non 
\linebreak
self-conjugate scalar particles with electrons (here seen to be the most 
\linebreak
relevant for determining the thermal decoupling epoch) 
is proportional to 
\linebreak
$(\fld \frd \, m_e^2) \, / \, (\mdmd \mFd)\,$. 
\ The decoupling is given by 
$\ a_{dec(dm-e)} = 5 \  10^{19} $ 
$\sigma_{el(dm-e)} \ c \ \tilde{n}_{dm}\ \rho_{e}/\rho_{dm}\ $ 
which is equivalent to $\ a_{dec(dm-e)} \ \thickapprox \ 5 \ 10^{19} $
$\sigma_{el(dm-e)} \ c \ \tilde{n}_e$ 
for $\,T_{dec(dm-e)} > \mdm\,$ and to 
$\ a_{dec(dm-e)}\, \thickapprox \,5 \ 10^{19} \ \sigma_{el(dm-e)} \ c \ \tilde{n}_e \ T_{dec(dm-e)}/\mdm$
for  $\,T_{dec(dm-e)} < \mdm\,$ (here $\tilde{n}_e$ is the comoving electron number density which drops down 
once the electrons become non relativistic). 
With the above cross section, we find a decoupling temperature $T_{nr} < $ MeV. 
However, because electrons annihilate around 500 keV, 
the thermal decoupling of \dm eventually occurs at a temperature close to $m_e$. 
Therefore, \dm particles with $\mdm> O(\UUNIT{MeV}{})$ should be able to 
annihilate so that their free-streaming scale should be indeed lower than 190 pc 
(corresponding to $\thickapprox 10 M_{\odot}$).

Thus, if light scalar \dm particles are coupled to neutrinos through neutral 
fermions, both the free-streaming scale (or, similarly in fact, the self-damping 
scale) and the neutrino-induced damping scale are negligible compared with the 
scale of primordial structures.

These conclusions would have to be softened if there were a close degeneracy 
between $\mdm$ and $m_F$ (\ie if $\,m_F^2-\mdmd  < 2 \ \mdm E_{\nu}$) 
since, in this case, the elastic scattering cross section becomes roughly 
constant\footnote{The denominator, proportional to $m_F^4$, is indeed 
replaced by $\mdm\!^2\,T^2$ which suppresses the dependence in $T^2$ in the numerator.}
and  one has to use 
the constraint given by eq.(\ref{dmnu}) instead of 
eq.(\ref{tdeux}). This appears more promising for large values 
of the \dm mass ($\mdm \!> O(100)$ GeV) because the case  
$\mF \sim \mdm$ cannot be excluded as yet, 
but this is, on the other hand, 
extremely unnatural because of the mass degeneracy it requires.  
Assuming, however, such a degeneracy (and $\fl =1$, $\mdm \sim 100$ GeV), 
one finds a damping mass greater than $1 \ M_{\odot}$  
but the required mass difference 
$\mF - \mdm $ is about a few hundred keV's!  
Without such a degeneracy, the linear P(k) associated with 
\dm particles of $\mdm \sim 
100$ GeV is expected to be damped much 
below $1\,  M_{\odot}$ (with $ \fl =1$ and taking $\mF \sim 100$ GeV), as 
the corresponding interactions decouple before $T \sim 1$ MeV.

Let us now consider the elastic scattering cross section of scalar 
\dm particles with neutrinos through the exchange of a $\U$ boson. 
We find 
$$ \sigma_{dm-\nu} \ c \ T_{dec(dm-\nu)}^{-2} \thickapprox \ 10^{-33} \ C_{\U}^2 \ f_{U_l}^2 \ 
\left(\frac{m_{\U}}{\UUNIT{MeV}{}}\right)^{-4} \UUNIT{cm}{5} \UUNIT{s}{-1}.$$

Using $C_U  f_{U_l} \simeq (3-12) \ 10^{-8} \ (\mdm/\UUNIT{MeV}{})^{-1} \  (m_U/\UUNIT{MeV}{})^2$, 
as imposed by relic density (and  $g-2$) constraints, 
we merely see that $\,b_{el(dm-\nu)}\,$ is about 
$\,b_{el(dm-\nu)} \,\thickapprox \,(1-15) \ 10^{-48} \ 
\left(\mdm/\UUNIT{MeV}{}\right)^{-2} \ \UUNIT{cm}{5} \UUNIT{s}{-1}\,.$ \,Thus, 
the neutrino-induced damping in the case of a light gauge boson 
(and scalar \dmsb) mainly depends on the \dm mass. By comparing (\ref{tdeux}) 
with the above expression of $\,b_{el(dm-\nu)}$, one finds that the damping scale 
associated with particles having a $\mdm \sim 4$ MeV for instance 
is about $3 \ 10^4 \, M_{\odot}$, if we assume that the coupling 
of the $U$ boson to neutrinos is the same as to electrons.
(More precise values are derived in 
\cite{bhds}.) 
It may be reasonable, however, to consider that the $U$ boson does not
couple to electrons and neutrinos with the same strength, especially in view 
of avoiding a too large contribution  
to the $\nu e$ elastic scattering cross section at low energy
(which requires typically $\,|f_{U\,\nu}\, f_{U\,e}/m_U^{\,2}|\,\simle\,G_F\,$),
while keeping a sufficiently large value for the annihilation cross
section of Dark Matter particles.
In particular, decreasing the coupling of the $U$ boson to neutrinos 
would then tend (depending on the $U$ coupling to Dark Matter $\,C_U\,$) to make
the damping mass even smaller than the above estimate.
This damping scale gets, also, obviously smaller for larger \dm masses.


To  conclude this section, let us mention that the calculation of 
the elastic scattering cross section of light \dm particles 
with neutrinos appears important when \dm is coupled to a $U$ boson 
because the damping in the linear power spectrum 
is large enough to be quoted here. However, even for $\mdm < m_{\mu}$, it is still  too small 
to modify the non linear matter power spectrum at cosmological scales. In fact, we 
expect that the smallest scales which are erased in the linear P(k) be 
regenerated once entered in the non linear regime; so 
the \textit{non linear} matter power spectrum at $z=0$ and $10^{4} M_{\odot}$ 
might eventually appear similar to that in Cold Dark Matter scenarios \cite{spergel,julien},  
despite the unusual \dm characteristics we are considering. 
In which case, it should be difficult to identify  
the nature of \dm particles from the non linear and linear matter power and CMB spectra only.

\ghost{ But the study of reionization might  help to distinguish them.  
On the other hand, the possibility of a very light \dm and a very light $U$ seems 
to have consequences at visible scales (the extreme case we considered, namely $\mdm \sim 3$ MeV 
implies a damping of $\thickapprox \ 10^9 M_{\odot}$) which means 
that they should be borderline to be excluded (still remembering that i) the damping 
mentioned here is in the linear matter power spectrum, and ii) a light \dm could perhaps 
also affect the reionization epoch, so one should have to be careful in using 
the constraints from the linear matter power spectrum)}

\subsection{Damping scales of spin 1/2 candidates} 

In the previous subsection, we found that the neutrino-induced 
damping scale of a scalar \dm coupled to a fermion $F$ was null 
when \dm is self-conjugate. We now estimate this collisional 
damping scale in the case of spin-1/2 \dm candidates.  Both 
the elastic scattering cross section of Majorana and Dirac spin-1/2 fermions 
are found to be proportional (in the local limit) to  
$\sigma_{\tiny{el(\mbox{dm}-\nu})} \propto \frac{\flq \ }{16 \ \pi \ \mFq} \ T^2 \ .$
\ghost{\\
\sigma_{\tiny{el(\mbox{dm}-\nu)}} &\sim& \frac{\flq \ }{4 \ \pi \ \mFq} \ T^2 \ \ \mbox{for dm$\ =\overline{\mbox{dm}}$} . 
\nonumber 
\end{eqnarray}}
As expected, these cross sections are temperature (or neutrino energy) 
dependent 
\,($\sigma_{el(\tiny{\mbox{dm}-\nu)}}\, c \simeq \,
b_{el(\tiny{\mbox{dm}-\nu})} \ T^2$) 
so that the associated damping scale is given in terms of the  parameter 
$b_{el}$ (see eq.\,(\ref{tdeux})). However, unlike scalars, 
the cross section for self-conjugate (\ie here Majorana) 
particles does not vanish. 

Because these cross sections are actually very close to that obtained with 
scalar \dmsb, the collisional damping effect in the linear matter power spectrum  
should be of the same order of magnitude than what is found for scalar candidates 
(say $M \le 10-0.3 M_{\odot}$) for $m_{F} \gtrsim 100$ GeV (and $\mdm \in [1-100]$ MeV 
respectively). 
Also, for heavy \dm particles (\ie  
$\mdm \sim 100$ GeV), one expects the damping scale to be 
negligible $M \ll 1 \  M_{\odot}$ 
unless there is an extremely close degeneracy, \ie a few keV, 
between $m_{F}$ and 
$\mdm$, in which case one gets a larger damping scale as a result of the enhancement 
of the neutrino-\dm elastic scattering cross section.  
(If such a degeneracy 
exists, a more detailed study is 
required in order to take into account the co-annihilations between 
$F$ and the \dm particles.)

\section{Case of an asymmetry}

To satisfy the relic density requirement, we searched 
for the conditions which give rise to a \dm annihilation cross section 
 of the order of 
$ \sigma_{ann} \ v_{rel} \ \sim 10^{-26}-10^{-27} \ 
(\Omega_{dm} h^2/0.1) \  \UUNIT{cm}{3} \UUNIT{s}{-1}$ in the 
unusual range $\mdm$ below a few GeV's. 
However, if there exists an asymmetry between the 
\dm particle and antiparticle number densities 
(which may be necessary for $\mdm \simle$ 100 MeV, in the case of $F$ exchanges),   
the annihilation cross section can be in fact  much larger 
than the value displayed previously (depending on the size of the asymmetry). 

The question is then to determine whether larger values of the cross section are indeed possible    
within reasonable particle physics assumptions. 
We have seen that if \dm was coupled to a new gauge boson, the mass $m_U$ 
and the couplings $C_U$ and $f_{U_l}, f_{U_r}$ 
should be small enough (for small values of $\mdm$) in order to get the proper value of    
the annihilation cross section. Larger values of $ \sigma_{ann} \ v_{rel}$ 
can be obtained by increasing the coupling $C_U$ or decreasing $m_U$ 
but this requires an appropriate choice of parameters.  
On the other hand,  scalar \dm particles can 
naturally have a quite large annihilation cross section 
if they have non chiral couplings to a heavy fermion $F$ 
but its associated $g-2$ contribution should be cancelled by introducing 
a new set of particles. Once this ``way out'' accepted, one can indeed consider larger values of the 
couplings and justify an asymmetry between $n_{dm}$ and $n_{dm^{\star}}$, 
that will allow to evade the gamma ray constraint 
(which is crucial for $\mdm \lesssim 100$ MeV).

For spin-1/2 Dirac fermions, on the other hand, 
it is rather difficult to consider such an asymmetry 
(unless $\mdm$ be larger than a few GeV) because that would require values of 
the Yukawa couplings $\fl$ and $\fr$ which could potentially make the theory 
non perturbative.

In any case, increasing the couplings allows for larger \dm elastic scattering cross sections 
with neutrinos and therefore larger neutrino-induced damping effects. However, if a spin-0 \dm particle is coupled 
a fermion $F$, the maximum damping effect should be definitely lower than 
a few ten $M_{\odot}$'s. 
If, on the other hand, \dm is coupled to a $U$ boson, 
with $C_U$ larger than the values displayed in Appendix, 
then the corresponding damping scale could be in principle significantly  
larger than what is found without any asymmetry, but 
a more detailled study is required as other damping effects may then be 
at work. Imposing the condition $\,l \lesssim 100$ kpc\, could then 
perhaps constrain  the maximum values of $C_U$ that is possible to consider.
\ghost{
Imposing the condition $\,l \lesssim 100$ kpc\, could then constrain  
the maximum values of $C_U$ that is possible to consider. but,  
presumably, smaller than a few $10^5 M_{\odot}$, which is still small compared to the 
presently observed cosmological scales. Also, 
it is worth remembering that this damping effect, expected in the linear 
matter power spectrum, may be absent in the non linear spectrum P(k) that one obtains 
at the present epoch \cite{spergel,julien}. }

\section{A possible candidate}

\vspace{-3mm}
We have determined the conditions that allow for 
light  scalar \dm candidates to satisfy both the relic density 
 and gamma ray constraints. They can be coupled to a 
heavy fermion (in which case an asymmetry between 
the particle and antiparticle number densities is required for $\mdm \lesssim 100$ MeV), or to a new 
gauge boson (associated with, at least, 
an extra $U(1)$ symmetry). 

\subsection{A new spin-0 gauge boson as a possible Dark Matter candidate}

Now, we would  like to point out that naturally stable \hbox{spin-0} 
Dark Matter candidates coupled to fermions 
may be already be present within an existing class of theories or models 
which display, 
at least partly, an underlying $N=2$ extended supersymmetry\,\cite{mirror}.
Or also, in higher-dimensional theories, whether supersymmetry 
is present or not\,\cite{6d}.
In both cases spin-1 gauge bosons are supplemented with new scalars which   
indeed appear as
{\it spin-0 ``gauge'' particles}, so that we now expect, in these frameworks, 
the existence of new spin-0 gluon octets, or spin-0 photons, for example.
The lightest of these spin-0 particles, if neutral and uncolored, may well be a viable 
Dark Matter candidate, along the lines discussed in the previous sections.


In such theories quarks and leptons are usually supplemented
with new {\it \,mirror partners}.
Indeed, in $N=2\,$ theories, quark and lepton fields belong to {\it \,matter
hypermultiplets\,} (the latter describing 4-component Dirac fermion fields,
i.e. left-handed as well as right-handed fields, both with the same gauge
symmetry properties). Left-handed electroweak doublets and right-handed singlets
associated with the usual quarks and leptons should then be accompanied
by right-handed doublets and left-handed singlets.
The latter would describe 
mirror partners ($F$) of the ordinary quarks and leptons ($f$).
Such mirror partners also tend to be naturally present in 
higher-dimensional theories \,-- whether or not they are supersymmetric --\,
since 4d Dirac fermion fields with, initially, vectorial couplings 
to the gauge fields naturally appear as a result of the dimensional 
reduction.


Since none of these mirror particles has been observed yet, 
neither at LEP nor at TEVATRON,
it is often believed that they simply don't exist at all,
or at least disappear from the ``low-energy'' spectrum as the result of an appropriate 
supersymmetry-breaking (or dimensional-reduction) mechanism.
We shall explore here the possibility according to which these mirror particles 
could indeed exist at not-too-high energies,
say of the order of a few hundred \,GeV$/c^2$\,'s, to fix the ideas.
In a more specific way mirror neutrinos
(with their right-handed component fields belonging to electroweak doublets)
are coupled to $Z$ bosons and 
should certainly be heavier than about 40 GeV$/c^2$,
\,otherwise they should have been
produced in $Z$ decays at LEP. Mass splittings within $SU(2)$ doublets of mirror particles 
also should not be too large, 
otherwise they could generate a too large contribution to the $\,\rho$ parameter of the 
electroweak theory.

Just as one expects, in ordinary supersymmetric theories, the existence of a
``lightest supersymmetric particle'' which may naturally be stable 
thanks to $R$-parity, one may also expect here the stability of some
``lightest mirror particle'', in relation with a $Z_2$ 
discrete symmetry $M_p$ under which mirror fermions (and sfermions, 
within supersymmetric theories) would be odd.
This symmetry may be referred to as ``mirror-parity'',
provided one carefully remembers that it does not exchange ordinary with mirror particles,
but simply changes the signs of the fields describing the latter.
New spin-0 gauge particles, which in these extended SUSY or higher-dimensional theories
couple ordinary to mirror fermions,
would then also be odd under this discrete $M_p$ symmetry operator.
The ``lightest $M$-odd particle'' (or LMP) is then expected to be absolutely stable,
if indeed $M$-parity is conserved, and a potential Dark Matter candidate.

This LMP could be a mirror neutrino
(in which case it should be heavy) or 
one of the new neutral spin-0 gauge bosons, which may well be light, 
the case of interest to us here.
We would then expect mirror particles to have decay modes 
such as:
\begin{equation}
\left\{\ 
\ba{ccc}
l_{\hbox{\tiny Mirror}}  &\to   &\ l\ +\ 
\hbox{new spin-0 gauge particle}\ \ ,  \vspace{.3cm} \\
q_{\hbox{\tiny Mirror}}  &\to& \ q\ +\ 
\hbox{new spin-0 gauge particle}\ \ .
\ea \right.
\end{equation}
The lightest of the new ``gauge scalars'' 
could then be absolutely stable,
if lighter than all mirror fermions.
It could be a spin-0 photon\,\footnote{Although 
a mirror neutrino is not directly coupled 
to a spin-0 photon, 
it could still decay into $\,\nu\,+$ spin-0 photon\, 
through radiative effects, or have other decay modes like 
$\,\nu_{\hbox{\tiny Mirror}}\,\to\,\nu\ e^+e^-\,+ $
spin-0 photon.
The lightest spin-0 mass eigenstate may also turn out 
to be a spin-0 partner of the weak hypercharge gauge field $B^\mu$, 
in which case it directly couples mirror neutrinos to ordinary ones.
} or, maybe more plausibly,
one of the spin-0 partners of the weak hypercharge
$U(1)$ gauge field $B^\mu$. 
(Note that these neutral gauge scalars, 
which have electroweak quantum numbers $\,T_3=Y=0$, are not coupled 
to the $Z$ boson so that they cannot be directly produced in $Z$ decays
at LEP.)


The pair production of unstable mirror particles 
would then ultimately lead to missing energy-momentum (precisely carried away by two
unobserved spin-0 photons or more generally neutral spin-0 gauge particles),
just as in supersymmetry, a signature of the pair production of SUSY particles
is missing energy-momentum carried away by two unobserved photinos 
or neutralinos, the ``lightest SUSY particle'' being stable.


It is also possible (at least in principle) to consider a continuous
$M$-symmetry
instead of a $M$-parity (related to $M_p$ by $\,M_p=(-1)^M$).
\,The corresponding lightest stable particle would then carry 
plus (or minus) one unit  of an additive
conserved quantum number $M$, \,and would be described
by a complex field  $\frac{a-ib}{\sqrt2}$.
\,This particle would then differ from its antiparticle, both being stable
but carrying opposite values
($\pm1$) of the conserved additive quantum number $M$.
\,This situation may be particularly interesting in the presence of an
initial
asymmetry between the corresponding numbers of particles and
antiparticles,
whether this asymmetry has been present from the beginning,
or induced by $M$-violating interactions at some point in the
evolution of the Universe.
The relic abundance of the lightest, stable, $\,M=+1\,$ particle may then
be
determined essentially by the size of the particle/antiparticle asymmetry
(provided one can achieve a sufficiently large value
of the annihilation cross-section),
as it is believed to happen
for the relic abundance of baryons in the Universe.
This is especially relevant for light \dm candidates,
which could both acquire the correct relic density while at the same time
\,-- in the presence of such an aymmetry --\,
avoiding the gamma ray constraint which applies to candidates lighter
than about 100 MeV,
as discussed in Section 7.
This would require, however, the above $M$-symmetry to survive unbroken
down to low energies, a requirement which tends
to be in conflict with
the necessity of generating masses for both ordinary fermions $f$
and mirror fermions\footnote{This concerns only a continous
$M$-symmetry,
not the discrete symmetry of $M$-parity,
which in any case governs the stability of the
\dm candidate considered here, whether or not they are their own
antiparticles.} $F$.


Finally, we also note that the chiral (or non chiral) character of the
couplings
of such new spin-0 \dm particles to fermion/mirror-fermion pairs is
generally related
with the complex (or real) character of the corresponding fields;
but that mixing effects associated
with mechanisms of mass generation and symmetry breaking may also generate
non chiral couplings from couplings which were, initially, of a chiral
nature.
This is essential in the phenomological analysis, as we saw earlier.
We should of course be attentive to the fact that the
requirement of non chiral couplings (in view of sufficiently large
annihilation cross sections allowing for a suitable value of the relic
abundance)
has to be reconciled with the constraints from the anomalous magnetic
moments
of charged leptons (as discussed in Appendix \ref{gm2a}).


\subsection{Extended supersymmetry}

Furthermore, in an extended SUSY framework (possibly although not necessarily 
associated with a higher-dimensional supersymmetric theory) 
we are led, through the consideration 
of two $Z_2$ discrete symmetries, to a natural framework 
for {\it two neutral stable particles}, e.g. a rather familiar neutralino LSP 
plus a spin-0 photon (or companion of the weak hypercharge gauge field $B^\mu$) LMP.
Depending on the values of their masses and interaction rates, 
they may turn out to be acceptable dark matter candidates, 
under the conditions discussed earlier.

Let us recall first some important features of $N\!=\!2$
extended supersymmetric theories\,\cite{mirror}.
When these are formulated with an $N\!=\!1$ superfield formalism, 
ordinary gauge superfields 
are accompanied by additional ($N\!=\!1$) ``chiral gauge superfields'', 
so as to describe, 
jointly, $\,N\!=\!2\,$ massless gauge multiplets.
At the same time matter (and also Higgs) chiral superfields systematically 
occur in pairs, so as to describe $\,N=2\,$ ``hypermultiplets''.
$\,N\!=\!2\,$ gauge multiplets describe in particular 
two gluino octets, two photinos, ... as well as two 
color-octets of spin-0 gluons, two spin-0 photons, etc.:
\bc
\small
$
\begin{array}{|c|c|c|}
\hline && \\  [-.2truecm]
\ \ \hbox{photon}\ \ & \ \ \hbox{2 spin-}\frac{1}{2}\ \hbox{photinos}\ \ & 
\ \ \hbox {2 spin-0 photons} \ \  \\  [-.2truecm] && \\
\hbox{gluons} & \ \hbox{2 spin-}\frac{1}{2}\ \hbox{gluino octets} \ & 
\ \hbox {2 spin-0 gluon octets}\
\\  [-.2truecm] && \\  \hline
\end{array}
$
\ec
\normalsize
These particles all remain massless, as long as the $N=2$ supersymmetry is kept unbroken.
These new ``scalar gauge  fields''
may also appear as originating 
from the fifth and sixth components  of higher-dimensional gauge fields 
$\,V^{\hat\mu}$, \,in a six-dimensional spacetime\,\cite{6d}.
Each pair of gaugino fields in 4 dimensions then originates from a single 8-component 
chiral gaugino field, in a 6-dimensional spacetime.

As we have seen extended supersymmetric theories, however, 
have the unpleasant feature of 
naturally leading to vectorlike and even vectorial theories,
necessitating the introduction of chirality-breaking mechanisms 
for matter fields, and of mechanisms responsible for quark and lepton masses. 
This led us to consider mirror lepton and quark fields.
Even if we are led, at a certain point, to abandon the full $N=2$ supersymmetry 
in favor of a simple $N=1$ subalgebra (or even of no supersymmetry at all), 
it is still conceivable that 
some sectors (e.g. gauge and Higgs bosons, together with their superpartners)
could display visible signs of the underlying extended supersymmetry.
Whether this is indeed the case or not, however, does not affect significantly
our present analysis of possible Dark Matter candidates.

The $SU(3)\times SU(2)\times U(1)\,$ gauge group
may be spontaneously broken into the $\,SU(3)\times U(1)\,$ subgroup of 
QCD $\times$ QED, with the help of \,-- at least --\, two electroweak Higgs 
doublet  fields, which are described in the $N=1$ superfield formalism by the familiar 
$H_1$ and $H_2$ chiral superfields,
as usual in the (M)SSM.
\,It is however more elegant to use a {\it \,quartet\,} of spin-0 Higgs 
doublets.
After the electroweak symmetry breaking the $\,W^\pm\,$ and $\,Z\,$ 
massive gauge bosons then belong 
to complete charged and neutral massive gauge multiplets of the extended 
$N=2$ supersymmetry. These multiplets also describe 4 Dirac winos 
(or $\!$ 4 Majorana zinos)
appearing as gaugino-higgsino mixtures, as well as 
5 charged (or 5 neutral) \,spin-0 Higgs bosons
(i.e. altogether 32 degrees of freedom for the massive gauge multiplet 
of the $W^\pm$, and 16 for the $Z$ boson) \cite{mirror}.
All these particles have the same mass $m_W$, or $m_Z = m_W/\cos\theta$, 
as long as the $N\!= 2$ supersymmetry is kept unbroken.
In the harmonic superspace formalism appropriate to the description of
$N=2$ extended supersymmetry, all four doublet Higgs fields may be described 
by a single $N=2$ electroweak doublet Higgs superfield 
\be
\omega=\left(\ba{c} \omega^0 \\ \omega^- \ea \right)\ \ .
\ee
Its charged component 
as well as the imaginary part of its neutral component 
get eliminated, or ``eaten away'', while the $\,N=2\,$ gauge superfields describing the 
$\,W^\pm\,$ and $\,Z\,$  acquire masses \cite{superfield}. 
The remaining uneaten real part 
of $\omega^0$ describes 4 neutral spin-0 Higgs bosons 
accompanied by two Majorana higgsinos.

\vskip .1truecm

Whether the extra spin-0 components of gauge fields should actually show up 
or not in the low-energy theory depends on the details of the mechanism 
that should be responsible for the breaking of the extended supersymmetry
(and/or the compactification of the extra space dimensions).
The breaking of the supersymmetry may be elegantly obtained by demanding periodic 
and antiperiodic boundary conditions for ordinary $R$-even particles and their $R$-odd 
superpartners, respectively -- in which case the masses of the (lowest-lying)
gravitinos, gluinos and photinos,
which fix the energy scale at which supersymmetric particles should start 
to show up, would be given, in the simplest case 
and up to radiative correction effects, by
\be
\label{comp}
m_{3/2}\  =\ m_{1/2}\ =\ \frac{\pi\,\hbar}{L\,c}\ =\ \frac{\hbar}{2\,R\ c}\ 
\ee
in terms of the size $L$ of the extra dimension responsible for supersymmetry 
breaking
(or of the corresponding ``radius'' $R$).
This led us to consider the possibility of relatively ``large'' extra dimensions,
associated with a compactification scale that could then be 
as ``low'' as $\,\sim $ TeV scale\,\cite{compac}.
In such a framework, widely discussed now, it is 
quite conceivable that the new spin-0 states,
as well as mirror lepton and quark fields, may only manifest themselves  
at the compactification scale. On the other hand, 
it is also legitimate to discuss their
physical effects, should they actually be present in the low-energy theory,
much below the compactification scale.

\ghost{
Whether the extra spin-0 components of gauge fields should actually show up 
or not in the low-energy theory depends on the details of the mechanism 
that should be responsible for the breaking of the extended supersymmetry
(and/or the compactification of the extra space dimensions).
While it is quite conceivable that these new spin-0 states,
as well as mirror lepton and quark fields, may only manifest themselves  
at the compactification scale (which, however, 
might be as ``low'' as $\sim$ TeV for example\cite{compac}), 
it is also legitimate to discuss their
physical effects, should they actually be present in the low-energy theory,
much below the compactification scale.
}

\subsection{Two discrete symmetries}

The new $M$-odd spin-0 states are expected to couple ordinary matter 
(lepton and quark) fields to {\it \,mirror\,} lepton and quark fields,
as discussed earlier.
We do not discuss here the mechanism by which ordinary 
as well as mirror lepton and quark fields should acquire their masses, 
which may lead us to abandon, at some point, the full $\,N=2\,$ supersymmetry.
Just as a $N=1$ supersymmetric theory may admit a discrete $\,R$-parity 
symmetry -- which is a discrete $\,Z_2\,$ remnant of a continuous 
$\,R$-symmetry, with $\,R_p=(-1)^R$  --\,  
an extended $\,N=2\,$ theory in general admits 
an extended $\,SU(2)\,$ or $\,SU(2)\,\times U(1)\,$  
global $R$-symmetry group, acting on the doublet of supersymmetry generators,
\begin{equation}
\left( \ba{c} Q^1_{\,L} \vspace{2mm} \\ Q^2_{\,L} 
\ea \right)\ \ .
\end{equation}

One can then initially define two distinct $R$-parity symmetry operators,
$R_{1\,p}$ and $R_{2\,p}$,
associated  with the first and second supersymmetry generators, respectively,
the usual $\,R$-parity symmetry then corresponding to the product
\begin{equation}
R_p\ =\ R_{1\,p} \ R_{2\,p}\ \ .
\end{equation}
The $R_1$ and $R_2$ parities within the $N=2$ massless gauge multiplets
and matter multiplets are 
given as $\,|R_{1\,p} \ R_{2\,p}\!\!>$ in the Table below,
together with the corresponding values of $\,R$-parity. 
By singling out one of the two supersymmetry generators (say $Q_1$), 
we may also redefine one of the two $R_{i\,p}$ discrete symmetries (say
$R_{2\,p}$) as the previously discussed mirror-parity symmetry $M_p$, 
a point to which we shall return soon.


\scriptsize
\bc
$
\ba{|c|c|c|}
\hline \hline && \\  [0truecm]
\hbox{gauge bosons:}& 
\,\hbox{2 spin-}\frac{1}{2}\ \hbox{gauginos:}\,& 
\,\ba{c} \hbox {2 spin-0}\ \ \ \ \ \ \ \ \ \ \  \\ 
\ \ \ \ \ \hbox{gauge scalars:} \ea\, \\ 
 && \\  [0truecm]  
|++>  &   \left\{\ \ba{c} |-+>  \vspace{5mm} \\ |+-> \ea \right. & 
|-->  
\\  [+.7truecm] 
\ R_p=+1,\ M_p=+1\
& R_p=-1,\ M_p=\pm1 & \ R_p=+1,\ M_p=-1 \ \\  [0truecm] && \\
\hline \hline &&  \\  [0truecm]
\ba{c} \hbox{leptons, quarks:}    \vspace{2mm} \\ \ l,\ q \,:\ \ |++>\ea & 
\left\{
\ba{c} \hbox{sleptons, squarks:}\vspace{.3cm} \\
\tilde l,\ \ \ \tilde q\ \ : \ \ |-+>\vspace{4mm}  \\
\hbox{\it ``smirrors''} \vspace{.3cm} \\
\widetilde l_{\hbox{\tiny M }},\ \widetilde q_{\hbox{\tiny M}}\!:|+-> 
\ea \right. & 
\ba{c} \hbox{\it mirror} \\ \!\hbox{leptons and quarks:}\!  \vspace{2mm} \\
l_{\hbox{\tiny M }},\,q_{\hbox{\tiny M }}:\ |-->  \ea  
\\  [+1.2truecm] 
\ R_p=+1,\ M_p=+1\
& R_p=-1,\ M_p=\pm1 & R_p=+1,\ M_p=-1 \\  [-.1truecm] && \\ [-.1truecm] && \\  \hline \hline
\ea
$
\ec
\normalsize

\vspace{2mm}

Among the four electroweak Higgs doublets which tend to be
naturally present 
in such theories, which all have $R$-parity equal to $+1$
\,(with $\,|R_{1\,p} \ R_{2\,p}\!\!> \ =\, 
\hbox{\scriptsize $|++>,\ |++>,\ |-->$}$ \ and\ 
$\hbox{\scriptsize $|-->$}$\,),
\,two, which acquire non-vahishing v.e.v.'s,
may be defined so as to have $M$-parity (identified as $R_{2\,p}$) equal to $+1$, 
while the two others (which keep vanishing v.e.v.'s) have $M$-parity equal to $-1$; 
the spontaneous breaking of the electroweak symmetry keeping intact the discrete 
mirror-parity symmetry, as well as the usual $R$-parity.

\vspace{-.1cm}

\subsection{Two stable Dark Matter candidates, from extended supersymmetry}

\vspace{-.4cm}

Once we have chosen to restrict the full $N=2$ supersymmetry of the 
initial theory to a simple and familiar $\,N=1\,$ supersymmetry, 
the $\,R_{2\,p}\,$ discrete $\,Z_2\,$ symmetry may also be identified 
to the ``mirror-parity'' operator 
that we denoted previously as $M_p$.
It commutes with the first SUSY generator ($Q_1$)  
and is equal to $+1$ for the ordinary particles of the 
supersymmetric standard model, including those described 
by the two doublet Higgs superfields 
$H_1$ and $H_2$. $M$-parity, on the other hand, is equal to $-1$
for the mirror particles, their superpartners, and for the spin-0 gauge particles 
which couple ordinary leptons and quarks to their mirror partners, 
together with the new inos (second octet of gluinos, 
additional charginos, neutralinos, and Higgs bosons)
necessitated by the original extended supersymmetry.

The first two Higgs doublets, with $M_p=+1$, 
correspond to the spin-0 components of the familiar 
$H_1$ and $H_2$ superfields
of the (M)SSM, responsible for lepton and quark masses.
The same Higgs doublets described by $H_1$ and $H_2$
could also be responsible 
for very heavy mirror lepton and quark masses (should such particles 
effectively be present in the low-energy theory),
if one assumes large Yukawa couplings to mirror particles, 
without generating a breaking of the discrete $M$-parity symmetry.
In this framework mirror quark and lepton masses \,-- which occur 
in violation of the $\,SU(2)\times U(1)\,$ 
electroweak gauge symmetry --\, are expected to be
smaller than a few hundred GeV/$c^2$'s \,(unless one is ready 
to tolerate very large values of Yukawa coupling constants, 
leading to mirror particles significantly heavier than the $\,W^\pm$ and $Z$'s).
Given the limited success, within ordinary supersymmetry, to find 
attractive and predictive models of supersymmetry-breaking,
we should leave open the range of possibilities 
for the masses of the new particles. This allows us, in particular,  to 
pay special phenomenological attention to the possibility 
of a new light stable spin-0 Dark Matter particle, as we have done here.

\section*{Conclusion}

\vspace{-.3cm}

In this paper, we discussed the possibility that scalar particles as light as a few MeV (and no
diagonal coupling to the $Z$ boson) could be a viable explanation to the 
\dm issue. In particular, we determined the 
conditions for the mass range [\,$O(\UUNIT{MeV}{})-O(\UUNIT{GeV}{})$\,]  
to simultaneously satisfy particle physics, relic density and 
gamma ray  constraints. We found that: 

\begin{itemize}
\item Scalar \dm  particles  coupled to  heavy fermions $F$ 
($\mF > 100$ GeV) can have an annihilation cross section 
into ordinary fermions large enough to give rise to the observed 
value of $\Omega_{dm} h^2$ but this requires  
the coupling dm-$\bar{f}$-$F$ to be non chiral.
However, when this condition is satisfied, 
the production of gamma rays associated with light particles 
($\mdm \lesssim 100$ MeV),  
in the energy range below one hundred MeV's, 
is found to be much larger than what is observed. 
Thus, in order to maintain light particles as a viable solution to the 
\dm problem, there should exist 
an asymmetry between their particle and antiparticle number densities. 
On top of that, one may also have to introduce additional particles 
(\eg charged scalars and neutral fermions or a pseudoscalar particle), to cancel  
the too large contribution $a_{\mu, e}^{\mF}$ that is required 
to get the desired value of the annihilation cross section.
The new particles we introduce are expected 
to be within the reach of future LHC experiments, so it 
 should be possible to test and maybe to exclude the possibility of 
light scalar \dm particles in a close future.

\vspace{2mm}
The range $\mdm > O(100 \UUNIT{MeV}{})$, on the other hand,  does not require 
any asymmetry assumption and appears 
much easier to test and presumably to rule out, 
notably through gamma ray predictions (if this asymmetry does not exist indeed) or 
from various experiments, like direct 
\dm searches (for $\mdm >$ GeV). 

\vspace{6mm}

\item  Scalar \dm  particles coupled to a new gauge boson $U$ 
have an annihilation cross section  naturally proportional to 
$v_{dm}^2$ at the freeze-out epoch. Therefore, the gamma ray constraint  
appears much easier to achieve for candidates with $\mdm \lesssim 100$ MeV. 
Since, on the other hand, their annihilation 
cross section turns out to depend significantly on the \dm mass 
(as in the case of fermionic \dmsb),  one can get an acceptable relic density provided 
the $U$ boson is light enough and the couplings sufficiently small 
(when \dm is light). Basically, to satisfy both the relic density 
requirement and the gamma ray constraints, one can take, for instance,  
$\mdm \sim 4$ MeV, 
$m_U \sim 10$ MeV, $C_U \sim 4 \ 10^{-3}$ and $f_{U_l} \sim 4 \ 10^{-4}$. 
These values, directly 
compatible with the electron and muon 
$\,g-2\,$ constraints, correspond to
a $U$ boson lifetime of the order of $\thickapprox \ 10^{-15}$ s. 

\vspace{2mm}
If coupled to a $U$ boson, 
these light \dm particles ($\mdm \lesssim 100$ MeV) should yield a gamma ray flux lower 
than 
$10^{-5} \left(\frac{\mdm}{\UUNIT{MeV}{}}\right)^{-2}$ cm$^{-2}$ s$^{-1}$, \,which seems close (for $\mdm = 1$ MeV)   
but still out of the reach of present gamma ray experiments for $\mdm > O(\UUNIT{MeV}{})$. 
However, a gammay ray signature from the galactic centre at low
energy could be due to the annihilation of such light Dark Matter particles.
And residual annihilations in structures at 
``large redshift''  could perhaps yield reionizing photons that might influence 
structure formation.  Thus light \dm particles coupled to a $U$ boson might  
be possible and quite interesting regarding their implications on structure formation. 
\end{itemize} 

In conclusion, we found two different scenarios
in which \dm  could be made of light scalar particles. 
The first one relies on the existence of heavy (charged) fermionic particles $F$ (\eg mirror fermions)  
and presumably other particles to cancel $g-2$ contributions,  
and the second one, on the existence of a light gauge boson. 
There might exist other scenarios that allow for light scalar candidates 
(\eg with a dominant coupling to quarks or couplings with a light neutrino of a new kind) 
but we 
did not explore them. To end, we mentioned a possible framework,
originating from extended supersymmetry and/or extra spacetime dimensions, 
in which scalar particles would appear 
as natural \dm candidates.

\section*{Acknowledgments}
The authors would like to thank T. Abel, J. Collins, J. Devriendt, K. Hamilton,
J. Iliopoulos,  J. Silk, D. Spergel, M. Strikman and J. Taylor 
for helpful discussions. 
C.B. is supported by an individual PPARC Fellowship.

\section{Appendix \label{sec:annex}} 

We now give the expressions of the squared  
matrix elements associated with the cross sections that enter the estimate of the relic density and 
damping scales of \dm candidates. We disregard the 
annihilations into two photons, expected to be negligible 
compared with annihilations into a pair 
fermion-antifermion. We also disregard the annihilations into 
$W^+ W^-$ or $Z Z$ only relevant for $\mdm \geq m_W, m_Z$ 
respectively.

We shall consider only tree-level processes where either a $F$ or 
$\U$ boson is exchanged, disregarding possible interference 
terms which are not expected to modify our conclusions.

\subsection{Annihilation cross sections associated with the 
exchange of a $F$ \label{anns}}

All expressions regarding \dm annihilations are given in terms of the 
inverse of the squares of the $t$-channel and $u$-channel denominators, 
developed as:  
$$ (t-m_F^2)^{-2} \ =\  T_0 + \ T_1 \ p_{dm} \ \cos{\theta} + \ p_{dm}^2 \ (T_{20} + T_{21} \cos{\theta}^2)\ ,$$ 
$$ (u-m_F^2)^{-2} \ =\  T_0 - \ T_1 \ p_{dm} \ \cos{\theta} + \ p_{dm}^2 \ (T_{20} + T_{21} \cos{\theta}^2)\ ,$$ and 
$$((t-m_F^2) \ (u-m_F^2))^{-1}\ =\  T_0 + \ p_{dm}^2 \ (T_{20} + T_{21} \cos{\theta}^2/3)\ ,$$
with 
\begin{eqnarray}   
T_0 \ \ &=& \ \ \frac{1}{(- \mFd - \mdmd + \mfd)^2}\ \ , 
\nonumber \\ \nonumber \\
T_1 \ \ &=&\ \ \frac{- 4 \ \sqrt{\mdmd - \mfd}}{(- \mFd - \mdmd + \mfd)^3}\ \ , 
\nonumber \\ \  \nonumber \\
T_{20}\   &=&\ \ \frac{4}{(-\mdmd + \mfd - \mFd)^3}\ \ , 
\nonumber \\ \nonumber \\
T_{21}\   &=&\ \ \frac{12 \ (\mdmd - \mfd)}{(\mdmd - \mfd + \mFd)^4}\ \ . 
\nonumber
\end{eqnarray}

In the following, we compute and display the quantity 
$$|{\mathcal{M}}|^2 \ = \ \int^{+1}_{-1} \ |M|^2 \ d \cos{\theta_{cm}}$$ 
where $|M|^2$ is estimated in the centre of mass reference frame. With these 
conventions, the annihilation cross section times velocity is related 
to $|{\mathcal{M}}|^2$ by:
$$ \sigma_{ann} \,  v_{rel} \ =\  \frac{1}{16 \ \pi \ s} \ 
\ \sum_{s_i, s_j} \ \frac{1}{(2 s_i +1) \, (2 s_j +1)} \  \ 
|{\mathcal{M}}|^2\ \ ,$$
$\sum_{s_i, s_j}$ denoting the usual sum over the spins of the incoming 
particles.  
Let us start with the annihilation of spin-1/2 Majorana fermions.

\subsubsection{\underline{Majorana \dmsb}}

\begin{center}
\hspace{-1cm}
\begin{picture}(80,80)(20,50)
\Line(50,110)(90,110)
\Line(50,70)(90,70) 
\ArrowLine(90,110)(130,110)
\ArrowLine(130,70)(90,70)
\DashLine(90,70)(90,110){5}
\Text(135,110)[]{f}
\Text(135,70)[]{$\bar{f}$}
\Text(40,110)[]{dm}
\Text(40,70)[]{dm}
\Text(82,90)[]{$F$}
\end{picture}
\hspace{2cm}
\begin{picture}(80,80)(20,50)
\Line(50,110)(90,110)
\Line(50,70)(90,70) 
\ArrowLine(90,70)(108,88)
\Line(112,92)(130,110)
\ArrowLine(130,70)(90,110)
\DashLine(90,70)(90,110){5}
\Text(135,110)[]{f}
\Text(135,70)[]{$\bar{f}$}
\Text(40,110)[]{dm}
\Text(40,70)[]{dm}
\Text(82,90)[]{$F$}
\end{picture}
\end{center}
\footnotesize
\begin{eqnarray}
\!\!\!\!\!\!\!\!\!\!\!\!\!
|{\mathcal{M}}|^2 &=&\ 
 8 \, T_0 \, \mdmd \, \bigg(\,2 \,\fl \fr \,\mdm + (\fld + \frd) \,\mf \,\bigg)^2 
\nonumber\\ &&\nonumber\\ &&
+ \ \frac{8 \ p_{dm}^2}{3} \  \Bigg[\ 
2 \, T_0 \    \bigg( 
2 \ \mdmd \ (\,2 \flq \,+\, 9 \fld \frd \,+ \, 2 \frq\,)  
\nonumber \\ && \ \ \ \ \ \ \
+ \ 6 \ \fl \, \fr \, (\fld + \frd) \ \mdm \ \mf  - \ 
(\flq \,+ \,12 \fld \frd \,+\,\frq) \ \mfd    \bigg)  
\nonumber\\ &&  
- \  4 \ T_1  \ 
 (\fld + \frd) \ \mdm \, \sqrt{\mdmd \!-\! \mfd}\,
\left((\fld \!+ \!\frd) \,\mdm + 2 \,\fl \fr \,\mf \right)
\nonumber \\ &&
 + \ 3 \ \mdmd \  T_{20} \ \
\bigg(2 \ \fl \fr \ \mdm \ + \  (\fld + \frd) \ mf \bigg)^2 
\nonumber \\ &&
 + \ \mdmd \ T_{21} \ \ \bigg( \, 4 \ (\flq + 3 \fld \frd + \frq) \ \mdmd 
\nonumber \\ && \ \ \ 
\ + \  
20 \,\fl \,\fr \ (\fld \!+\! \frd)  \,\mdm \,\mf + 
(\flq + 18\,\fld \frd + \frq) \,\mfd \, \bigg) \, / \, 3 \ \Bigg]. \nonumber \\   
 \nonumber
\end{eqnarray}

\normalsize
Our expression is consistent with that displayed 
in \cite{ko} for Bino-like particles. In particular, one readily sees 
that the S-wave annihilation cross section of 
two Majorana spin-1/2 particles into an ordinary fermion-antifermion 
pair is suppressed proportionately to $m_f^2/\mFq$ (in the local limit approximation) 
in the case of chiral couplings ($\fl \fr =0$) while 
both the $v_{dm}$ dependent and independent terms 
turn out to be  comparable if one allows for  
non chiral couplings. This is not expected to bring any drastic changes 
in the case of a light \dm particle ($\mdm < O(\UUNIT{GeV}{})$) because of the small mass difference 
between $\mdm$ and $\mf$ but this may be more interesting in supersymmetry,
when the lightest neutralino is heavier than the top quark ($m_{\chi} > m_t$) and 
there exists a mixing angle between the two stop eigenstates ($\tilde{t}_l$ and 
$\tilde{t}_r$).

\subsubsection{\underline{Dirac \dmsb}}
\begin{center}
\hspace{-1cm}
\begin{picture}(80,80)(20,50)
\ArrowLine(50,110)(90,110)
\ArrowLine(90,70)(50,70) 
\ArrowLine(90,110)(130,110)
\ArrowLine(130,70)(90,70)
\DashLine(90,70)(90,110){5}
\Text(135,110)[]{f}
\Text(135,70)[]{$\bar{f}$}
\Text(40,110)[]{dm}
\Text(40,70)[]{$\overline{dm}$}
\Text(82,90)[]{$F$}
\end{picture}
\end{center}
In the case of spin-1/2 Dirac fermions, we find: 

\footnotesize

\begin{eqnarray} 
\!\!\!\!\!\!\!\!\!\!\!\!
|\mathcal{M}|^2 \ &=&\ 
8 \ T_0 \ \mdmd \ 
\bigg(\, (\fld + \frd) \, \mdm + \ 2 \, \fl \, \fr \, \mf \bigg)^2 
\nonumber \\&& \nonumber \\&& \!\!\!\!\!\!\!\!
+\  \frac{8 \  p_{dm}^2}{3} \ \ \Bigg[  
\ T_0 \ (\fld + \frd) \  
\nonumber \\ && \ \ \ \ \ \ \ \ \ \ \ \ \  
\bigg( 7 \ (\fld \!+\! \frd)  \,\mdmd + \,12\,\fl \fr \, \mdm \, \mf - 
(\fld \!+\! \frd) \, \mfd   \bigg)   
\nonumber \\ && \!\!\!\!\!\!\!\!
- \ 2 \ T_1 \  
 \mdm \, (\fld + \frd) \ 
\sqrt{\mdmd - \mfd} \ \bigg((\fld + \frd) \,\mdm \,+ \,2 \,\fl \fr \, 
\mf\bigg) 
\nonumber \\ && \!\!\!\!\!\!\!\!
+ \ (3 \ T_{20} + T_{21})
 \ \mdmd \ 
\bigg((\fld + \frd) \, \mdm  + \, 2 \ \fl \ \fr \mf\bigg)^2 \ 
 \Bigg] \ . \nonumber \\   \nonumber 
\end{eqnarray} 
\normalsize
Both the S and P-wave contributions (or more precisely the velocity dependent and independent terms) 
are now dominated by $\mdmd/\mFq$ (in the local limit) when $\fl \fr =0$. This is in contrast with spin-1/2 
Majorana fermions where the S-wave is found to be proportional to $\mfd/\mFq$ (in the same situation), 
indicating that Dirac and Majorana spin-1/2 particles do not behave in 
the same way.

\subsubsection{\underline{Self-conjugate scalar \dmsb} \label{sec:annscal}}
\begin{center}
\hspace{-1cm}
\begin{picture}(80,80)(20,50)
\DashLine(50,110)(90,110){5}
\DashLine(50,70)(90,70){5}
\ArrowLine(90,110)(130,110)
\ArrowLine(130,70)(90,70)
\Line(90,70)(90,110)
\Text(135,110)[]{f}
\Text(135,70)[]{$\bar{f}$}
\Text(40,110)[]{dm}
\Text(40,70)[]{dm}
\Text(82,90)[]{$F$}
\end{picture}
\hspace{2cm}
\begin{picture}(80,80)(20,50)
\DashLine(50,110)(90,110){5}
\DashLine(50,70)(90,70){5}
\ArrowLine(90,70)(108,88)
\Line(112,92)(130,110)
\ArrowLine(130,70)(90,110)
\Line(90,70)(90,110)
\Text(135,110)[]{f}
\Text(135,70)[]{$\bar{f}$}
\Text(40,110)[]{dm}
\Text(40,70)[]{dm}
\Text(82,90)[]{$F$}
\end{picture}
\end{center}
In the case of the annihilation of   
two self-conjugate scalars, one finds:
%
\footnotesize
\begin{eqnarray} 
\!\!\!\!\!\!
|\mathcal{M}|^2  \ &=&\                                 
16 \ T_0 \ \,\big( \, (\fld + \frd)  \, \mf + 2 \, \fl \, \fr \, \mF \big)^2 \ \,
(\mdmd \, - \, \mfd) 
\nonumber\\  && \nonumber\\  &&
+ \ \frac{16}{9} \ p_{dm}^2 \ 
\bigg( \, (\fld + \frd)\, \mf + \,2 \, \fl \fr \, \mF \bigg) 
\nonumber\\  &&
\!\!\!\!\!\!\!\!\!\!\!\!\!\!\!\!\!\!\!
\Bigg[ \ 9 \ \bigg( (\fld + \frd)  \mf + 2 \fl \fr \mF \bigg) \, T_0 - 
 3 \ T_1 \ (\fld + \frd) \,\mf \ \sqrt{\mdmd - \mfd}
\nonumber\\  &&
\!\!\!\!\!\!
+ \ 
(9 \, T_{20} \ + \ 2 \, T_{21}) 
\ \bigg( \, (\fld + \frd) \, \mf \ + \ 2 \fl \, \fr \, \mF \, \bigg) \ 
(\mdmd - \mfd)  \ \Bigg] \ . \nonumber 
\end{eqnarray}

\normalsize
The S-wave contribution is seen to be proportional to $1/\mFd$ when the couplings are 
non chiral (instead of $\mfd/\mFq$ as obtained for Majorana and $\mdmd/\mFq$ for Dirac particles). 
The corresponding cross section then appears almost independent of the \dm mass, suggesting that 
it should be possible to get the correct relic density for light \dm particles. 


\subsubsection{\underline{Non self-conjugate scalar \dmsb} \label{annscni}}
\begin{center}
\hspace{-1cm}
\begin{picture}(80,80)(20,50)
\DashLine(50,110)(90,110){5}
\DashLine(50,70)(90,70){5} 
\ArrowLine(90,110)(130,110)
\ArrowLine(130,70)(90,70)
\Line(90,70)(90,110)
\Text(135,110)[]{f}
\Text(135,70)[]{$\bar{f}$}
\Text(40,110)[]{dm}
\Text(40,70)[]{dm$^{\star}$}
\Text(100,90)[]{$F$}
\end{picture}
\end{center}
The squared amplitude associated with the annihilation of non 
self-conjugate scalars into a fermion and antifermion pair 
is given by:


\footnotesize
\begin{eqnarray}   
\!\!\!\!\!\!\!\!\!\!\!\!\!\!\!\!\!\!\!\!                                                               
|\mathcal{M}|^2 \ &=&\  \ 4 \ T_0 \ \,(\mdmd - \mfd) \ \,
\bigg((\fld + \frd) \mf + 2 \fl \fr \mF \bigg)^2  
\nonumber \\ && \nonumber \\ && \ 
+ \ \frac{4 p_{dm}^2}{3} \ 
\Bigg[ - \bigg( \ (\fld + \frd) \mf + 2 \fl \fr \mF \bigg)  
\nonumber \\ && \nonumber \\ &&
\!\!\!\!\!\!\!\!\!\!\!\!\!\!\!\!
\bigg(2 \,T_1 \ (\fld \!+\! \frd) \,\mf \ \sqrt{\mdmd - \mfd}
 \ + 
\nonumber \\ && \ \ \ \ \ \ 
\ (3 \ T_{20} + T_{21}) \  (-\mdmd + \mfd)
\ \large( \ (\fld \!+\! \frd) \mf + 2 \fl \fr \mF \large)    \bigg) 
\nonumber\\ && \nonumber\\ && 
\!\!\!\!\!\!\!\!\!\!\!\!\!\!\!\!
 +  \ 2 \ T_0 \ \bigg(   6 \,\fl \, \fr \, (\fld +\frd) \ \mf \ \mF 
\, + \, (\flq+\frq)  \ (2 \mdmd + \mfd) \ 
\nonumber\\ && 
\ \ \ \ \ \ \ \ \ \ \ \ \ \ \ \ \ \ \ \ \ \ \ \ \ \ \ \ \ \ \ \ \ \ \ \ 
\ \ \ \ \ \ 
\ \ \ \ \ \ \ \ \ \ \ \ 
+ \ 6 \ \fld \ \frd \ (\mfd + \mFd)  \ \bigg) \  
\Bigg]\ .  \nonumber \\   \nonumber
\end{eqnarray}
\normalsize
Here, both the S and P-wave contributions are proportional to $1/\mFd$ in the case of non chiral couplings. 
Whether the couplings are chiral or not, our expressions for self-conjugate and non self-conjugate scalars 
are similar so the assumption $dm = dm^*$ or $dm \neq dm^*$ is not crucial concerning the derivation of the 
annihilation cross section of scalar particles.

\subsection{Annihilation cross sections associated with the 
production of a $\U$ boson}
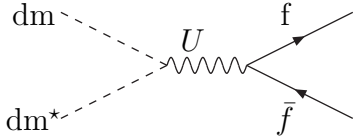
\begin{center}
\hspace{-3cm}
\begin{picture}(80,80)(20,50)
\DashLine(50,110)(90,90){3}
\DashLine(50,70)(90,90){3} 
\ArrowLine(120,90)(160,110)
\ArrowLine(160,70)(120,90)
\Photon(90,90)(120,90){3}{5}
\Text(135,110)[]{f}
\Text(135,70)[]{$\bar{f}$}
\Text(40,110)[]{dm}
\Text(40,70)[]{dm$^\star$}
\Text(100,100)[]{$U$}
\end{picture}
\end{center}
The cross section corresponding to 
the virtual $s$-channel production of a $\U$ boson is seen to be given by:
\footnotesize
\begin{eqnarray} 
\!\!\!\!
|\mathcal{M}|^2 \  &=&\    
\frac{16 \ C_{\U}^2 \ p_{dm}^2}{3  \ (m_{\U}^2 - 4 \, \mdmd)^2} 
\ \left[4 \mdmd (f_{\Ul}^2 + f_{\Ur}^2) - \mfd (f_{\Ul}^2 - 6 f_{\Ul} f_{\Ur} + f_{\Ur}^2) 
\right]  \ .    \nonumber                         
\end{eqnarray}
\normalsize
This cross section is very small when the 
$\U$ boson is very heavy but it is, on the other hand, of the good order of magnitude 
when the $U$ boson is light enough, thereby allowing for light \dm to satisfy 
the relic density condition (in which case, the $U$ boson should be very weakly coupled). 

\subsection{Elastic scattering cross sections of \dm with neutrinos \label{nusc}}

Let us now derive the square of the matrix element associated with 
the elastic scattering cross section 
of \dm particles with neutrinos. Because the 
damping effects are expected to be due to free-streaming neutrinos 
(\ie which have already decoupled), 
we shall estimate the corresponding cross sections at a time 
$t>t_{dec(\nu)}$ (where $t_{dec(\nu)}$ is 
the neutrino thermal ``decoupling'' time). Since we are dealing with 
\dm particles heavier than a few MeV's, the condition $t>t_{dec(\nu)}$ 
can be translated in terms of the neutrino energy as: $E_{\nu} < \mdm$. 
Any dependence in $E_{\nu}$ would therefore 
tend to suppress the cross section.

\subsubsection*{\underline{Scalar \dm particles }}

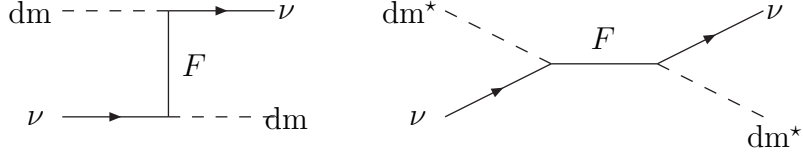
\begin{center}
\hspace{-2.5cm}
\begin{picture}(80,80)(20,50)
\DashLine(50,110)(90,110){5}
\ArrowLine(50,70)(90,70) 
\ArrowLine(90,110)(130,110)
\DashLine(90,70)(130,70){5}
\Line(90,70)(90,110)
\Text(135,110)[]{$\nu$}
\Text(135,70)[]{dm}
\Text(40,110)[]{dm }
\Text(40,70)[]{$\nu$}
\Text(100,90)[]{$F$}
\end{picture}
\hspace{2cm}
\begin{picture}(80,80)(20,50)
\DashLine(50,110)(90,90){5}
\ArrowLine(50,70)(90,90)
\ArrowLine(130,90)(170,110)
\DashLine(170,70)(130,90){5}
\Line(90,90)(130,90)
\Text(175,110)[]{$\nu$}
\Text(175,63)[]{dm$^\star$}
\Text(37,110)[]{dm$^\star$}
\Text(40,70)[]{$\nu$}
\Text(110,100)[]{$F$}
\end{picture}
\end{center}

The matrix elements corresponding to the elastic scattering of scalar \dm 
particles with ordinary neutrinos through a fermion $F$ exchange are given by: 
\begin{eqnarray}
M_{u \, ; \, el} \ &=& \ i \, \ubar{\nu_2} \, \fl P_R \, 
(\dslash{p}{\nu_2} - \dslash{p}{dm_1} + m_F) \, \fl P_L \, 
\u{\nu_1}/(u-\mFd)\ ,\nonumber \\                
M_{s \, ; \, el} \ &=&\  i \, \ubar{\nu_2} \, \fl P_R \,
(\dslash{p}{\nu_1} + \dslash{p}{dm_1} + m_F) \, \fl P_L  \, 
\u{\nu_1}/(s-\mFd)\ .\nonumber 
\end{eqnarray}
The $s$-channel does not contribute to the elastic scattering of 
 ``non self-conju\-gate'' \dm particles. We make 
the reasonable assumption of chiral couplings $dm-\bar{\nu}-F^0$ 
(which, within conventions defined in Section \ref{dmcoupling}, 
means $\fr = 0$). This assumption actually kills the 
contribution in $m_F$ in the numerator, thereby preventing the elastic 
cross section to be ``very'' large. 
From the matrix elements given above, we found:
\begin{eqnarray}
 |M|^2_{dm\neq dm^{\star}; \, el} \ \ &=&\ \
\frac{4 \, \flq}{ (t-\mFd)^2} \ \,
\bigg[\, (p_{dm_1}.p_{\nu_1})\, (p_{dm_1}.p_{\nu_2}) \,-\, 
\frac{\mdmd}{2} \ (p_{\nu_1}.p_{\nu_2}) \,\bigg] \nonumber \\     
&=&\ \ \frac{ \flq}{ (u-\mFd)^2} \ \,
\bigg( - s u + \mdmq \bigg)\ \ ,
\nonumber \\                                               
&&\hspace{-1cm} \mbox{and} 
\nonumber \\ 
 |M|^2_{dm = dm^{\star};  \, el} \ \ &=&\ \ 0\ \ \ \ \ 
\mbox{in the local limit} \, , \nonumber 
\end{eqnarray}
The reason why 
$ |M|^2_{dm = dm^{\star}; el}$ vanishes identically  in the local limit 
can be understood by rewriting the two matrix elements $M_t$, 
$M_u$ with the use of the Dirac equation. The latter reduce 
(in the local limit approximation) to:   
\begin{eqnarray}
M_u \ \,&=&\,\ - \,\fld \, \ 
\ubar{\nu_2} \,  \dslash{p}{dm_1} \, P_L \, \, \u{\nu_1}\,/\mFd \ ,\nonumber \\                
M_s \ \,&=&\,\ \ \fld \, \
\ubar{\nu_2} \, \dslash{p}{dm_1} \, P_L \, \, \u{\nu_1}\,/\mFd \ ,\nonumber 
\end{eqnarray}
so we find indeed $M_u + M_s =0$. \,A similar cancellation ($M_u + M_s=0$) 
also occurs when computing 
the annihilation process of two self-conjugate scalar \dm particles 
in a pair neutrino-antineutrino. This can be checked by 
replacing $m_f$ and $\fr$ by zero in the results found in
Section \ref{sec:annscal}.
On top of that, a partial wave analysis indicates that 
each of these matrix elements should be proportional to the neutrino energy 
so that the cross section for $dm \neq dm^{\star}$
is indeed expected to be suppressed proportionately to 
the square of the neutrino energy.  

Let us now estimate the contribution of a new gauge boson. 

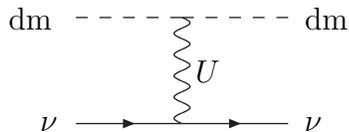
\begin{center}
\hspace{-3cm}
\begin{picture}(80,80)(20,50)
\DashLine(50,110)(90,110){5}
\ArrowLine(50,70)(90,70) 
\DashLine(90,110)(130,110){5}
\ArrowLine(90,70)(130,70)
\Photon(90,70)(90,110){3}{5}
\Text(145,110)[]{dm}
\Text(140,70)[]{$\nu$}
\Text(35,110)[]{dm }
\Text(40,70)[]{$\nu$}
\Text(100,90)[]{$U$}
\end{picture}
\end{center}

The only possible diagram is a $t$-channel. 
The corresponding matrix element can be written as: 
\begin{eqnarray}
M_t \ &=&\ Q^{\nu} \ C_U \, \fzlp \, 
\bigg(\,-g_{\mu \nu} + \frac{k_{\mu} k_{\nu}}{m_U^2}\,\bigg) \ \,
\frac{\ubar{\nu_2} \gamma^{\mu}  P_L \,  \u{\nu_1}}{(t-m_U^2)} \ ,
\nonumber 
\end{eqnarray}
where $\,Q^{\nu} = (p_{dm_1} + p_{dm_2})^{\nu}\,$ and 
$\,k_{\nu}= (-p_{dm_1} + p_{dm_2})_{\nu}$. 
\,This provides: 
$$
|M_t|^2 \ =\  \frac{C_U^2\, \fzlp^2}{(t-m_U^2)^2}  \  
\bigg( \, (s - u)^2  + t \ (4 \mdmd \! - t) \bigg) $$
which is proportional to the square of the neutrino energy.

\subsubsection*{\underline{Fermionic \dm particles}}

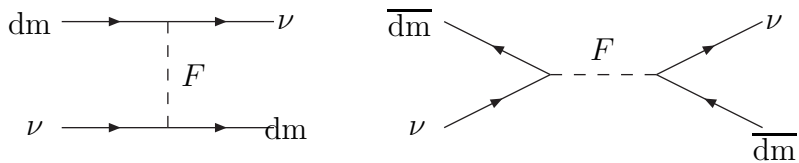
\begin{center}
\hspace{-3.5cm}
\begin{picture}(80,80)(20,50)
\ArrowLine(50,110)(90,110)
\ArrowLine(50,70)(90,70) 
\ArrowLine(90,110)(130,110)
\ArrowLine(90,70)(130,70)
\DashLine(90,70)(90,110){5}
\Text(135,110)[]{$\nu$}
\Text(135,70)[]{dm}
\Text(40,110)[]{dm }
\Text(40,70)[]{$\nu$}
\Text(100,90)[]{$F$}
\end{picture}
\hspace{2cm}
\begin{picture}(80,80)(20,50)
\ArrowLine(90,90)(50,110)
\ArrowLine(50,70)(90,90)
\ArrowLine(130,90)(170,110)
\ArrowLine(170,70)(130,90)
\DashLine(90,90)(130,90){5}
\Text(175,110)[]{$\nu$}
\Text(175,63)[]{$\overline{\mbox{dm}}$}
\Text(37,110)[]{$\overline{\mbox{dm}}$}
\Text(40,70)[]{$\nu$}
\Text(110,100)[]{$F$}
\end{picture}
\end{center}
The matrix elements corresponding to the elastic 
scattering of fermions on neutrinos (disregarding the $\U$ contribution) 
are given by:  
\begin{eqnarray}
M_{u \, ; \, el}\ \ &=&\ \ \ubar{\nu_2} \, \fl P_R \, \u{dm_1}\ \
\frac{1}{u - m_F^2} \ \ \ubar{dm_2}  \fl P_L \, \u{\nu_1}\ , \nonumber \\                
M_{s \, ; \, el}\ \ &=&\ \ \ubar{\nu_2} \, \fl P_R \, \v{dm_2} \ \
\frac{1}{s - m_F^2} \ \  \vbar{dm_1} \, \fl P_L  \, \u{\nu_1}\ .\nonumber 
\end{eqnarray}
Once again, the $s$-channel is expected to only contribute to the cross section associated with Majorana 
particles (but the $s$-channel actually also contributes for Dirac particles  which would 
interact with antineutrinos). We found:  
\begin{eqnarray}
 |M|^2_{dm\neq \overline{dm }} \ \ &=&\ \  
4 \ \fl^4 \,\  (p_{dm_1}.p_{\nu_2}) \, (p_{dm_2}.p_{\nu_1})\,/\, (u-\mFd)^2 
\nonumber \\                                               
&=&\ \ \flq \ \,\frac{(u-\mdmd)^2}{(u-\mFd)^2} \ \ ,
\nonumber \\                                               
&&\hspace{-1.6cm} \mbox{and} 
\nonumber \\ 
 |M|^2_{dm = \overline{dm }} \ \ &=&\ \  \flq \ 
\bigg(\,\frac{(u-\mdmd)^2}{(u-\mFd)^2}
 \ + \ \frac{(s-\mdmd)^2}{(s-\mFd)^2}\, \bigg)\ \ .
 \nonumber
\end{eqnarray}

Unlike scalars, there is no cancellation when 
$dm = \overline{dm}$ so that 
both Majorana and Dirac candidates are expected to  
yield a collisional damping effect, whether this effect turns out to be 
physically relevant or not. 
Here, we do not estimate the contribution of a new gauge boson (although
potentially useful) since fermionic \dm is not the main interest of this paper.

\subsection{Constraints from $g-2$ \label{gm2a}}

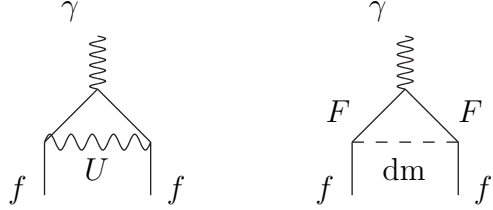
\begin{center}
\hspace{-1.5cm}
\begin{picture}(80,100)(20,50)
\Photon(90,130)(90,110){3}{5}
\Line(70,90)(90,110)
\Line(110,90)(90,110) 
\Photon(70,90)(110,90){3}{5}
\Text(80,140)[]{$\gamma$}
\Text(120,72)[]{$f$}
\Text(60,72)[]{$f$}
\Text(90,80)[]{$U$}
\Line(70,90)(70,70)
\Line(110,90)(110,70)
\end{picture}
\hspace{1cm}
\begin{picture}(80,100)(20,50)
\Photon(90,130)(90,110){3}{5}
\Line(70,90)(90,110)
\Line(110,90)(90,110) 
\DashLine(70,90)(110,90){5}
\Text(80,140)[]{$\gamma$}
\Text(120,72)[]{$f$}
\Text(60,72)[]{$f$}
\Text(115,102)[]{$F$}
\Text(65,102)[]{$F$}
\Text(90,80)[]{dm}
\Line(70,90)(70,70)
\Line(110,90)(110,70)
\end{picture}
\end{center}

One of the most stringent constraints on the existence of a light gauge boson 
and light \dm particles coupled to new fermions $F$ 
comes from the muon and electron anomalous magnetic moments. 
The corresponding contributions can be derived from \cite{leveille}.  


{\large $\bullet$}\ \ The diagram associated with the possible existence 
of a new neutral gauge boson $U$
is very similar to the $Z$ contribution.

\begin{table}
\vspace{3mm}
\caption{We display the expressions of the coupling ($f_{U_l}=f_{U_r}$) of the $U$ boson to the muon 
($U-\mu-\bar{\mu}$) and electron ($U-e-\bar{e}$) in terms of $\delta a_{\mu }$ and $\delta a_{e}$, 
the extra contributions of the $U$ boson to the muon and electron anomalous magnetic moments.
We also give the value of $C_U$, the $U$ boson coupling to the Dark Matter, that we 
obtain by using the relic density condition \ie by imposing that 
$\langle \sigma_{ann} v_{rel} \rangle_{ann}$ associated with the  virtual production of a $U$ boson be equal 
to $ 10^{-27}-10^{-26} \UUNIT{cm}{3} \UUNIT{s}{-1}$. $C_U$ is then given 
for both $10^{-27}$ and $10^{-26} \UUNIT{cm}{3} \UUNIT{s}{-1}$. We assume  
$\mdm \gtrsim 2 \mf$, $m_U \gtrsim 2 \mdm$.}  
\vspace{3mm}
\scriptsize
\begin{tabular}{c}
$
\ba{|c||c|c||c|c|} 
\hline  &&&&\\[-3mm]
m_U  &\delta a_{\mu}  & 
\ U \ \hbox{coupled to}\,\mu:\,f_{U_l}\,
& \delta a_e  & U \ \hbox{coupled to}\,e:\,f_{U_l} \\ [1mm]
\ \ (\,> 2 \ \mdm) \ \ &&\ U \ \hbox{coupled to}\ \mbox{dm}:\ C_U \
& &  \ U\ \hbox{coupled to}\ \mbox{dm}:\ C_U \ \\ [2mm] \hline 
&&&&\\
\mdm > m_{\mu} &
\  \simeq \, \frac{f_{U_l}^2}{12 \ \pi^2} \ \frac{m_{\mu}^2}{m_U^2}\ 
& \ f_{U_l} \simeq 3\, 10^{-6} \ \ \ \ \ \ \ \ \ \ \ \ 
&
\ba{c}\vspace{3cm} \\ 
\ \simeq \, \frac{f_{U_l}^2}{12 \ \pi^2} \ \frac{m_{e}^2}{m_U^2} \ 
\vspace{-3cm} 
\ea
&
\ba{c}
\, f_{U_l} \simeq 7 \ 10^{-5} \ \ \ \ \ \ \ \ \ \ \  \vspace{3mm} \\ 
\ \ \ \ (\frac{m_U}{\UUNIT{MeV}{}}) \, (\frac{\delta a_{e}}{10^{-11}})^{1/2} \ \\ \\
C_{U} \simeq \ (0.5-2) \ 10^{-3} \,\\ 
\ \ \ \ (\frac{m_U}{\mdm}) \, \left(\frac{\delta a_{e}}{10^{-11}}\right)^{-1/2} \ \\
\vspace{-7cm}\\
\ea
\\
&& \ \ \ \ (\frac{m_U}{\UUNIT{MeV}{}}) \, (\frac{\delta a_{\mu}}{10^{-9}})^{1/2} \ &&
\\ 
  &&&&\\
&& C_{U} \simeq (1-4) \ 10^{-2}\ \ \ \ \ \  && \\  
&&\ \ \ \ \ \ (\frac{m_U}{\mdm}) \,(\frac{\delta a_{\mu}}{10^{-9}})^{-1/2} \ &&\\ 
&&&&\\ \cline{1-3} 
&&&&\\
\mdm < m_\mu & \, \simeq\,  \frac{f_{U_l}^2}{8 \ \pi^2} \ & 
f_{U_l} \simeq 3 \, 10^{-4} \, 
(\frac{\delta a_{\mu}}{10^{-9}})^{1/2} \ &&\\
&&&&\\[-1mm]
&&&&\\
&&[ \ \hbox{\footnotesize no annihilation} &&\\
 &&\mbox{\small \ into}\ \mu^+\mu^-\ \hbox{\small possible}\ ] &&\\
&&&&\\
\hline
\ea
$
\label{table}
\end{tabular}
\end{table}

When $ m_{\mu}> m_U > m_e$, the  contribution of the $U$ boson to the anomalous 
magnetic moment of the muon is given by  
$\,\delta a_{\mu}^{m_{\mu} > m_{U}} \,\simeq\,  \frac{1}{8 \ \pi^2} 
(\,f_v^2 - f_a^2 \ C \ (\frac{m_{\mu}}{ M_U})^2 \,)$ 
where $\,f_v = (\fzlp +\fzrp)/2$, $\,f_a = (\fzlp - \fzrp)/2\,$ and $C$ is
a numerical coefficient. In order to 
avoid a potentially large (negative) contribution from $f_a$, and also for simplicity,
we shall assume $\fzlp =\fzrp$, 
i.e. a vector coupling of the new gauge boson.  
This may be naturally obtained by using
one extra Higgs singlet, in addition to the standard Higgs doublet,
to trigger the spontaneous breaking of the
$\, SU(3)\times SU(2)\times U(1)\times U(1)\,$ gauge group 
into the $\, SU(3)\times U(1)\,$ subgroup of QCD $\times$ QED \cite{bsp},
as discussed in subsection \ref{subsec:vec}.

This implies  
$\delta a_{\mu}^{m_{\mu} > m_{U}} \simeq 1.3 \ 10^{-2} \ f_{U_l}^2$,
that we have to compare with the  difference between the
experimental and standard model value, say (1-3) $10^{-9}$ 
(extendable up to 3 or 5  $10^{-9}$)
\cite{eedata}. From this comparison, 
one gets that $f_{U_l}$ or $\fzrp$, associated with the vector coupling 
$U-\mu-\bar{\mu}$ should be lower than about $(3-5) \ 10^{-4}$. 
Similarly, imposing that the contribution to the anomalous magnetic moment of the electron 
\,($\delta a_e^{m_{U} > m_e} \,\simeq \,\frac{f_{U_l}^2}{12 \ \pi^2} \ 
\frac{m_{e}^2}{m_U^2}$)\, be 
smaller than a few $10^{-11}$ (up to 8 $10^{-11}$),  
we find that $f_{U_l} (=\!\fzrp)$ associated with the coupling $U-e-\bar{e}$  
should be smaller than $\,(7-20) \ 10^{-5} \ (\frac{m_{U}}{\UUNIT{MeV}{}}) \,$, \,which gives 
at most $(7-20) \ 10^{-3} \,$ when $m_{\mu} > m_U > m_e$.

The coupling $C_U$ can be obtained by using the relic density argument, once 
the coupling $f_{U_l}$ fixed by the electron 
$g-2$ (and potentially by the muon $g-2$, 
if one assumes a universal relationship between the couplings of \dm to the muon and electron). 
Note that the study of the muon $g-2$ does not allow to derive 
the value of $C_{U}$ for $m_{\mu} > m_U$ since \dm (assume to be 
lighter than the $U$ boson) cannot annihilate into muons. 
Finally, for $m_{U} \gtrsim 10$ MeV, $\delta a_e = 3 \ 10^{-12}$, we find 
$f_{U_l}  =\fzrp \simle \ 4\ 10^{-4}$ and $C_U \simeq 4 \ 10^{-3}$.
One can do the same exercise for $m_U> m_{\mu}$; the corresponding 
values can be obtained by using the Table.

{\large $\bullet$}\ \ 
A new (charged) fermionic particle ($F$) will contribute to the $g-2$ of  
the muon and electron (hereafter denoted $l$) thanks to a diagram  which 
involves two $F$ particles and one \dm particle. 
Using \cite{leveille}, the contribution of the new $F$ particles 
is given by     
\bea
\delta a_{l}\ \  &\simeq & \ \ 
\frac{\ m_{l}^2}{32 \,\pi^2} \ \int_0^1\, dx \ \
\ba{c}
\frac{ (\fl + \fr)^2 \ (x^2 - x^3 + x^2 \frac{m_{F}}{m_{l}}) \ + \ 
(\fl - \fr)^2 \ (x^2 - x^3 - x^2 \frac{m_{F}}{m_{l}})}
{m_{l}^2 x^2 \ + \ (\mFd - m_{l}^2) x + \mdmd (1-x)}\ \ .
\ea
\nonumber
\eea
\normalsize
When $ \,m_{F} \gg m_{l}\,$ and $C_l\,C_r\neq 0$ (which is the case 
we are interesting 
\linebreak
in)  
the contribution 
$\,\delta a_{l}^{m_{F} \gg m_{l}} \simeq
\frac{\fl \ \fr \ m_{l}}{16 \pi^2 \, m_{F}}$\, 
is found to be proportional to $m_{l}/m_F$ \, rather than to $(m_{l}/m_F)^2\,$. 
\,To satisfy the $g-2$ constraints, the product $\fl \fr$ should then be 
about $ \,10^{-3}\, (m_F/100 \UUNIT{GeV}{})$ (or smaller). This is actually in 
contradiction with the relic density condition which imposes 
$\fl \fr \sim 10^{-2} (m_F/100 \UUNIT{GeV}{})$.

It therefore seems impossible to satisfy these two conditions simultaneously. However,   
other particles could bring a similar contribution to $g-2$ but with an opposite sign, thus cancelling the $F$ 
contribution. This can be achieved, for instance, by introducing a neutral fermion $F^0$, a 
charged scalar $H$ with a mass $\mF \sim m_{F^0} \gg m_H \gg \mdm$ and a coupling 
$F^0$-$H$-$l$  (somewhat equivalent in supersymmetry to $\chi^0$-$\tilde{l}$-$l$) 
of the same order as $F$-dm-$l$. The additional contribution to $g-2$ (associated with 
the diagram with two $H$ and a $F^0$) 
would be given (for $m_{F^0} \gg m_H$)
by $a_{l}^{H} \sim  - \frac{\fl_H \ \fr_H \ m_{l}}{16\, \pi^2 \, m_{F^0}} 
= - \frac{\fl_H \ \fr_H}{\fl \ \fr} a_{l}^{F}$. 
Therefore the contribution $a_{l}^{H}$ could actually cancel that of $a_{l}^{F}$ if one requires  
$\fl_H \,\fr_H \simeq  \fl \fr$ and some fine tuning between the $F$ and $F^0$ masses 
(basically one can use for instance 
$m_H \gtrsim O(100)$ GeV, $m_{F^0} \sim 10 \ m_H$, and $m_{F^-} \sim 11.8 \ m_H$). 
Such a spectrum might lead us to consider a multiplet including a neutral scalar 
\dm particle together with heavy charged scalar.

Such a cancellation can also occur quite naturally 
in the framework of theories originating from $N=2$ extended
supersymmetry or extra dimensions (cf. Section 8), through 
the introduction of  a new light neutral spin-0 particle. 
Both \dm and the new neutral spin-0 particle would have 
non chiral couplings but one would be a scalar and the other 
one, a pseudoscalar.  Their respective contributions to the muon or electron $g-2$ 
are expected to be of the same magnitude but of opposite sign. 
Their sum would be therefore naturally small (of the order of $(m_{\mu}/m_F)^2$), whether each individual contribution 
is large or not. This actually allows for a sufficiently large value of the 
annihilation cross section without being in conflict with $g-2$ constraints.

In a $N=2$ supersymmetric framework for example, both {\it \,scalar\,} and 
{\it \,pseudo\-scalar\,}
spin-0 photons would have, separately, non chiral couplings
to muons/ mirror muon (or electron/mirror electrons) pairs.
The non chiral character of the couplings
of a relatively light scalar spin-0 photon ($a_\gamma$),
here considered as a possible \dm candidate,
would allow for sufficiently
large annihilation cross sections ($a_\gamma a_\gamma\to f\bar f$).
The pseudoscalar spin-0 photon field ($b_\gamma$)
could have a somewhat larger mass (of e.g. a few GeV's, for example).
The dominant contributions of the $a_\gamma$ and $b_\gamma$ exchanges
to the charged lepton $g-2$ 's would then largely cancel out.
This may be easily understood since the complex combination
$(a_\gamma-i b_\gamma) /\sqrt 2$ behaves (as far as $g-2$ is concerned
and as long as the mass splitting between the two spin-0 photons is not
too large
compared to the mirror fermion mass $m_F$),
as a single complex field with chiral couplings,
resulting globally in a small (positive) contribution to
$\,a_\mu, e\, \propto \ \frac{\alpha}{6\,\pi}\ m_{\mu, e}^2/m_F^2\,$ (while 
the contribution is $\propto  m_{\mu, e}/m_F$ for $a_{\gamma}$ and $b_{\gamma}$ 
separately),
which is lower than about $10^{-9}$ for $m_F\ge 100$ GeV.


\subsection{$\U$ decay}

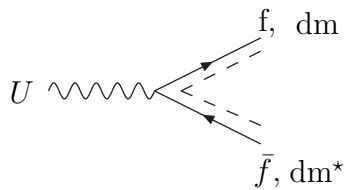
\begin{center}
\hspace{-2cm}
\begin{picture}(80,80)(20,50)
\Photon(50,110)(90,110){3}{5}
\ArrowLine(90,110)(130,130)
\ArrowLine(130,90)(90,110) 
\DashLine(100,110)(130,125){5}
\DashLine(130,97)(100,110){5} 
\Text(40,110)[]{$U$}
\Text(135,136)[]{f, }
\Text(135,80)[]{$\bar{f}$, }
\Text(152,136)[]{dm}
\Text(152,80)[]{dm$^{\star}$}
\end{picture}
\end{center}

The new gauge boson is supposed to be massive enough to decay  
into a pair fermion-antifermion (\eg $\,e^+ e^-$ and of course $\,\nu \bar{\nu}$),
and, if there is enough phase space, into two \dm particles.
The two body decay rate into a fermion-antifermion pair
is given by $\,\Gamma_U^{f} \simeq \frac{m_U \ f_{Ul}^2}{12 \pi }$, 
(assuming vectorial couplings $f_{Ul}=f_{Ur}$ for massive fermions), 
and the corresponding partial lifetime by 
$$\tau_{U}^{f} \ \sim\ 2.5 \ 10^{-12}\ (m_U/\UUNIT{MeV}{})^{-1} \, 
(f_{Ul}/10^{-4})^{-2}  \  \rm{s}\ .$$ 
\ghost{Using the couplings displayed in \ref{gm2a}, we find 
$\tau_{U}^{e} \ \sim \ 5 \ 10^{-12} (\delta a_e / 10^{-11})^{-1}\ 
(m_U/\UUNIT{MeV}{})^{-3}$ 
and, for $m_U > m_{\mu}$, $\tau_{U}^{\mu} \sim 2 \ 10^{-9} (\delta a_{\mu} / 10^{-9})^{-1} (m_U/\UUNIT{MeV}{})^{-3}$.}
The two body decay rate into two spin-0 (non self-conjugate) \dm particles is, on the other hand, 
given by 
$\ \Gamma_U^{dm}\simeq\frac{m_U \ C_{U}^2}{48 \pi } 
\simeq \frac{\Gamma_U^{f} \, C_{U}^2}{ (4 \ f_{Ul}^2)}\,$, 
which implies  
$$\tau_{U}^{dm} \ \sim \ 10^{-11} \ (m_U/\UUNIT{MeV}{})^{-1} \  
(C_{U}/10^{-4})^{-2}  \ \rm{s}\,. $$ 
With $C_{U} = 3 \ 10^{-3}$, as obtained if $\delta a_e = 3 \ 10^{-12}$, 
$\mdm= 4$ MeV and $m_U/\mdm = 2.5$, 
one gets $\,\tau_{U}^{dm} \sim 10^{-15}$ s. 
\,The corresponding value of $f_{U_l}$ for these parameters being of 
$4 \ 10^{-4}$, 
we find $\,\tau_{U}^{e} \sim 1.5 \ 10^{-14}  \, \rm{s}.$ 
In such situations the decay into two \dm particles 
can dominate over the decay into a pair fermion-antifermion 
(with a branching ratio $B_f = \Gamma_f/\Gamma_{tot}$ of about $6 \ 10^{-2}$).  
The $U$ boson would then mainly decay into ``missing energy''  
and would, most likely, also escape searches 
in nuclear transitions \cite{deboer}. This therefore indicates that, for some
values of the parameters, 
the coupling of the $U$ boson with \dm can allow for a short $U$ boson lifetime 
without violating the $g-2$ constraints (\ie without 
requiring a cancellation mechanism so as to satisfy 
the $g-2$ constraints)!

\subsection{Constraints from direct $U$ production and initial state radiation
}

The existence of a $U$ boson could also be constrained, directly or indirectly, 
by accelerator searches, through the processes $e^+ e^-
\rightarrow  \gamma \ U$ or $e^+ e^- \rightarrow U
\rightarrow$ dm dm.
Once produced directly (\ie $e^+ e^-
\rightarrow  \gamma \ U$), the $U$ boson may have invisible 
and visible decay modes (depending on whether it decays into 
dm dm, $\nu \, \bar{\nu}$ or $e^+e^-$, for instance). 
The invisible modes are expected to be dominant, according to 
the  branching ratios discussed previously, but the visible modes 
could nevertheless be important depending on the cross section associated with the $U$ boson 
production.

\subsubsection*{Direct $U$ boson production}
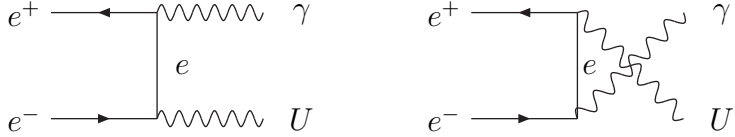
\begin{center}
\hspace{-3.5cm}
\begin{picture}(80,80)(20,50)
\ArrowLine(90,110)(50,110)
\ArrowLine(50,70)(90,70) 
\Photon(90,110)(130,110){3}{6}
\Photon(90,70)(130,70){3}{6}
\Line(90,70)(90,110)
\Text(145,110)[]{$\gamma$}
\Text(145,70)[]{$U$}
\Text(40,110)[]{$e^+ $}
\Text(40,70)[]{$e^-$}
\Text(100,90)[]{$e$}
\end{picture}
\hspace{2.5cm}
\begin{picture}(80,80)(20,50)
\ArrowLine(90,110)(50,110)
\ArrowLine(50,70)(90,70) 
\Photon(90,110)(130,70){3}{6}
\Photon(90,70)(130,110){3}{6}
\Line(90,70)(90,110)
\Text(145,110)[]{$\gamma$}
\Text(145,70)[]{$U$}
\Text(40,110)[]{$e^+ $}
\Text(40,70)[]{$e^-$}
\Text(95,90)[]{$e$}
\end{picture}
\end{center}
This process is similar
to $e^+ e^-$ annihilation 
into two photons  (especially for $\sqrt{s} > 10$ MeV)  in which one would replace a 
photon by a $U$ boson. One therefore expects the associated 
differential cross section $(\frac{d \sigma}{d \Omega})_{U \gamma}$ 
to be, at energy $E_e > m_U$, equal to 
$\left(\frac{d \sigma}{d \Omega} \right)_{U \gamma} \simeq 
2 \ (f_{U_l}^2 / e^2)\ 
(\frac{d \sigma}{d \Omega})_{\gamma \gamma}\,$, 
\,say  $\frac{ e^2 \, f_{U_l}^2 \ }{s} \ \frac{1+\cos \theta}{ \sin^2\theta}$, 
in the centre-of-mass frame (with $s = 4 E_e^2$). 
The $U$ boson's production should therefore 
be suppressed by a factor $ \simle 10^{-4}$ as compared to the annihilation 
into two photons. 
With the value $f_{U_l}  \sim 7 \ 10^{-5} \ \frac{m_U}{\UUNIT{MeV}{}} \ 
(\frac{\delta a_e}{10^{-11}})^{1/2}$ as mentioned earlier, 
we get a suppression factor of about $5 \ 10^{-8} \ (\frac{m_U}{\UUNIT{MeV}{}})^2 \ 
(\frac{\delta a_e}{10^{-11}})$. 
This implies a cross section smaller than 
$$\left(\frac{d \sigma}{d \Omega}\right)_{U \gamma}\  \simeq\ 5 \ 10^{-37} 
\ \left(\frac{E_e}{1 \UUNIT{GeV}{}}\right)^{-2} \ 
\left(\frac{f_{u_l}}{4 \ 10^{-4}}\right)^2 \ \frac{1+\cos \theta}{\sin^2\theta} 
\ \UUNIT{cm}{2}\ ,$$ 
which seems large enough  to be of interest for collider experiments depending on $f_{U_l}$. 
However, even in the case of a large cross section, 
very energetic photons, as produced in this process,   
might be easily confused with that originating from QED events (especially if the $U$ then decays in visible modes) 
so it might be difficult to exclude such a possibility.

If the $U$ boson mainly decays into \dmsb, then the $U$ production process turns out to be 
of the type $e^+ e^- \rightarrow \gamma + \dslash{E}{} \,$, 
where $\dslash{E}{}$ is missing energy, which  is of interest in experiments searching 
for single photon production events. 
But, in the case of a light \dm candidate, such a process is likely to remain unobserved, 
owing to the large background associated 
with $\,e^+e^- \to \gamma\,\gamma,$ in which one of the two photons
escapes detection.

Still it may be useful to compare our estimate to the sensitivity of 
``neutrino counting'' experiments, 
where two neutrinos are produced in $e^+ e^-$ annihilations and a single photon is used 
to trigger on the event ($e^+ e^- \rightarrow \nu \bar{\nu} \gamma$). 
The standard model cross section, at $\sqrt{s} = 29$ GeV,
for a tag photon with an energy greater than 1 GeV 
in the direction $\theta_{\gamma} > 20^{o}$, is about 0.04 pb 
for $N_{\nu}=3$ (and greater for $N_{\nu}>3$). 
The experimental constraint, at this energy, 
is $N_{\nu} < 7.9$ at 90$\%$ CL \cite{hearty}, which corresponds to a cross section of 
0.07 pb. Therefore extra particles with a cross section lower than 0.03 pb at $\sqrt{s} = 29$ GeV are, 
in principle, allowed.  
At LEP, where the limit on $N_{\nu}$ is 
much better because the energy in the centre-of-mass frame 
is closer to the $Z$ pole \cite{nulep,L3}, 
we expect the process $e^+ e^- \rightarrow U \gamma$ to be of the order of 
$\sigma \sim 5 \ 10^{-41} \left(\frac{f_{u_l}}{4 \ 10^{-4}}\right)^2 \ \UUNIT{cm}{2}$, 
say much lower than the cross section $e^+ e^- \rightarrow \nu \bar{\nu} \gamma$. This again
indicates that the existence of a $U$ boson should have escaped searches 
for ``missing energy''.

\subsubsection*{Anomalous single photon production}

Here are the diagrams involved in Initial State Radiation (ISR) mechanism: 

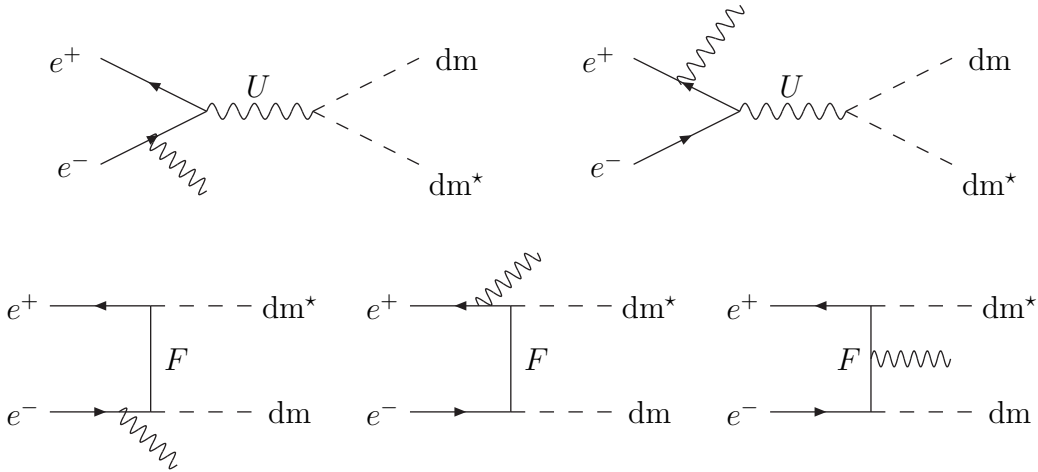
\begin{center}
\hspace{-3.5cm}
\begin{picture}(80,80)(20,50)
\ArrowLine(90,90)(50,110)
\ArrowLine(50,70)(90,90)
\DashLine(130,90)(170,110){5}
\DashLine(170,70)(130,90){5}
\Photon(90,90)(130,90){3}{5}
\Photon(68,80)(90,60){3}{6}
\Text(185,110)[]{dm}
\Text(185,63)[]{dm$^\star$}
\Text(37,110)[]{$e^{+}$}
\Text(40,70)[]{$e^-$}
\Text(110,100)[]{$U$}
\end{picture}
\hspace{4cm}
\begin{picture}(80,80)(20,50)
\ArrowLine(90,90)(50,110)
\ArrowLine(50,70)(90,90)
\DashLine(130,90)(170,110){5}
\DashLine(170,70)(130,90){5}
\Photon(90,90)(130,90){3}{5}
\Photon(68,100)(90,130){3}{6}
\Text(185,110)[]{dm}
\Text(185,63)[]{dm$^\star$}
\Text(37,110)[]{$e^{+}$}
\Text(40,70)[]{$e^-$}
\Text(110,100)[]{$U$}
\end{picture}
\end{center}

\begin{center}
\hspace{-2.5cm}
\begin{picture}(80,80)(20,50)
\ArrowLine(90,110)(52,110)
\ArrowLine(52,70)(90,70) 
\DashLine(90,110)(128,110){5}
\DashLine(90,70)(128,70){5}
\Line(90,70)(90,110)
\Photon(78,70)(100,50){3}{6}
\Text(143,110)[]{dm$^\star$}
\Text(143,70)[]{dm}
\Text(42,110)[]{$e^+ $}
\Text(42,70)[]{$e^-$}
\Text(100,90)[]{$F$}
\end{picture}
\hspace{1.7cm}
\begin{picture}(80,80)(20,50)
\ArrowLine(90,110)(52,110)
\ArrowLine(52,70)(90,70) 
\DashLine(90,110)(128,110){5}
\DashLine(90,70)(128,70){5}
\Line(90,70)(90,110)
\Photon(78,110)(100,130){3}{6}
\Text(143,110)[]{dm$^\star$}
\Text(143,70)[]{dm}
\Text(42,110)[]{$e^+ $}
\Text(42,70)[]{$e^-$}
\Text(100,90)[]{$F$}
\end{picture}
\hspace{1.7cm}
\begin{picture}(80,80)(20,50)
\ArrowLine(90,110)(52,110)
\ArrowLine(52,70)(90,70) 
\DashLine(90,110)(128,110){5}
\DashLine(90,70)(128,70){5}
\Line(90,70)(90,110)
\Photon(90,90)(120,90){3}{6}
\Text(143,110)[]{dm$^\star$}
\Text(143,70)[]{dm}
\Text(42,110)[]{$e^+ $}
\Text(42,70)[]{$e^-$}
\Text(82,90)[]{$F$}
\end{picture}
\end{center}

Because the direct production of \dm particles from 
$e^+ e^-$ annihilations consists only in ``missing energy'' (carried away 
by the two unobserved \dm particles), a possible 
signature to search for in accelerator experiments consists in the 
emission, by the incoming particles (or by the exchanged particles when one deals 
with a charged $F$), of a {\,\it single photon}: 
$$
e^+\ e^-\ \ \longrightarrow\ \ \gamma\ \ \mbox{dm} \ \mbox{dm} \ .
$$
This process is similar to the one involved in neutrino, neutralino or
sneutrino searches so we can use existing limits to determine 
whether light \dm particles can be viable or not.

The diagrams are basically the same as for sneutrino production 
but the $Z$ boson is replaced by a $U$ boson and 
the chargino  by a $F$ particle\,\cite{hearty}. 
The annihilations through the virtual production of a $U$ and a $Z$ boson seem
 at a first glance 
similar (the smallness of the couplings compensates the 
difference of mass between the $Z$ and $U$ bosons). However, because 
we are considering a light gauge boson, the cross section -- 
usually proportional to $G_F^2 E^2/ 12 \pi  \sim 10^{-39} (E/\UUNIT{GeV}{})^{2} \UUNIT{cm}{2}$ 
for $E < M_Z$ for ordinary weak interactions
--  is  
replaced by $\, \,f_{U_l}^2 C_U^2/E^2 \lesssim 6 \ 10^{-42} \left(\frac{m_U}{\UUNIT{MeV}{}}\right)^4  \ 
\left(\frac{\mdm}{\UUNIT{MeV}{}}\right)^{-2} \ (E/\UUNIT{GeV}{})^{-2} \UUNIT{cm}{2}$. 
This indicates that the pair production of \dm particles 
through the virtual production of a $U$-boson is indeed generally 
lower than the weak-interaction production of neutrino pairs. 
Anomalous single photon production due to 
$\,e^+ e^- \rightarrow  \mbox{dm} \, \mbox{dm}\, \gamma\,$ 
would then have escaped even ``low energy'' experiments 
(in particular, those appropriate for neutrino 
counting like PETRA and PEP experiments), 
if they proceed through the virtual production of a $U$ boson.

The \dm production through $F$ exchanges seems, on the other hand, much
closer to experimental limits, and   
therefore more interesting for accelerator experiments. 
The matrix elements associated with the emission of a photon by the 
incoming particles are given by: 
\small
\begin{eqnarray}
\!\!\!\!\!\!\!\!\!\!\!\!\!
M_1 \ &=&\ \frac{e}{D_1} \ \bar{v_1} \, (\fl P_R + \fr P_L) \, 
(\dslash{p_F} {} +  m_F)  \, 
(\fl P_L + \fr P_R) \, 
(\dslash{p_2} \, - \, \dslash{ k} \, + m_e)  \, \dslash{\epsilon} \,  u_2 \ ,
\nonumber \\
\!\!\!\!\!\!\!\!\!\!
M_2 \ &=& \ \frac{e}{D_2} \ \bar{v_1} \, 
\dslash{\epsilon} \, (\dslash{p_1} \, - \, \dslash{ k} \ + m_e)  
\, (\fl P_R + \fr P_L) \, (\dslash{p_F} {}+ m_F)  \, (\fl P_L + \fr P_R) \, 
u_2 \ , \nonumber \\ 
\ \label{matu}
\end{eqnarray}  
\normalsize
while the matrix amplitude associated with the emission of a photon by the exchanged 
particle can be written as: 
\small
\bea
\!\!\!\!\!\!\!\!\!\!
\label{mt}
M_3 = \frac{e}{D_3} \ \bar{v_1}  \ (\fl P_R + \fr P_L) \, 
(\dslash{p_{F_1}} {}+ m_F)  \,\dslash{\epsilon} \, 
(\dslash{p_{F_2}} {}+m_F)  \,  (\fl P_L + \fr P_R) \,  u_2\ ,
\nonumber \\ 
\eea
\normalsize
with $\,p_{F_1} = p_{e_1} - p_{dm_1}$, $\,p_{F_2} = p_{e_2} - p_{dm_2}$, 
$\,D_1 = 2 k.p_2 \ (p_F^2 - m_F^2)$, $\,D_2 = 2 k.p_1 \ (p_F^2 - m_F^2)\,$ and 
$\,D_3 = ((p_{e_1} - p_{dm_1})^2 - m_F^2) \ ((p_{e_2} - p_{dm_2})^2 - m_F^2)$.
The leading term in each squared matrix amplitude associated with (\ref{matu})   
is given, in the case of chiral couplings and neglecting the electron mass, 
by  $\,8 (\fl \fr)^2/ \mF^2$. The leading term in the squared matrix amplitude associated with 
(\ref{mt}) is, on the other hand, given by $\,4 \ E_e^2 \ (\flq + \frq)/\mFq$ for $E_e < m_F$.  
\,Since we consider values of $m_F$ greater than 100 GeV, the squared matrix amplitude 
associated with (\ref{mt}) appears slightly suppressed compared to that associated with  
(\ref{matu}). We can therefore neglect (\ref{mt}) to get a first estimate. 
 
Owing to the relation $\fld \frd/\mF^2 \thickapprox 4 \, \pi \, \sigma_{ann} v_{rel}$ 
(where we impose $ \sigma_{ann} v_{rel} \gtrsim 10^{-26}-10^{-27} \UUNIT{cm}{3} \UUNIT{s}{-1}$ 
so as to justify an asymmetry, as necessary if $\mdm \lesssim 100$ MeV), the single photon production\footnote{The cross 
section actually also depends on a term inversely proportional to $x$ 
but it is weighted by the square of the electron mass and is therefore suppressed.} cross section 
$e^+ e^- \rightarrow \mbox{dm} \, \mbox{dm} \, \gamma$ is expected to be of the order of 
$\,\sigma^{\gamma} \sim 4 \ \alpha \ x \ \langle \sigma_{ann} \, v_{rel} \rangle / 
4 \pi \ c$, \ie $ \,\sigma^{\gamma} \,\thickapprox x \ \frac{\langle \sigma_{ann} \, v_{rel} \rangle}{
10^{-26} \UUNIT{cm}{3} \UUNIT{s}{-1}} \ \UUNIT{fb}{}$, 
($x$ being defined as $E_{\gamma} = x \ E_e$ and assuming $E_e \gg \mdm > m_e$). 
This order of magnitude (although close enough to experimental limits 
to justify further studies since $\sigma^{\gamma} \sim \mathcal{L}_{\tiny{\mbox{LEP}}}^{-1}$), tends to   
suggest that light \dm particles could have indeed escaped previous 
collider (\eg PEP, LEP) experiments. (We note moreover that $\sigma^{\gamma}$ 
is also very close to the (LSP) neutralino-associated single-photon process for neutralino masses 
larger than present experimental limits.)

\subsubsection*{$U$-strahlung.}  

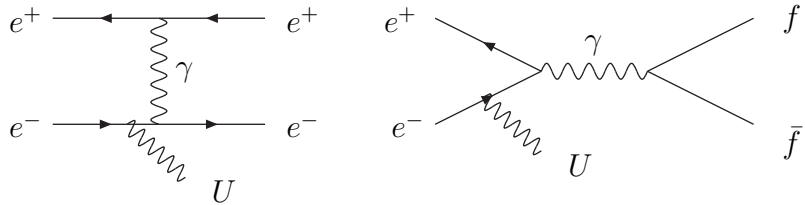
\begin{center}
\hspace{-3.5cm}
\begin{picture}(80,80)(20,50)
\ArrowLine(90,110)(50,110)
\ArrowLine(50,70)(90,70) 
\ArrowLine(130,110)(90,110)
\ArrowLine(90,70)(130,70)
\Photon(90,70)(90,110){3}{6}
\Photon(78,70)(100,50){3}{6}
\Text(145,110)[]{$e^+$}
\Text(145,70)[]{$e^-$}
\Text(40,110)[]{$e^+ $}
\Text(40,70)[]{$e^-$}
\Text(100,90)[]{$\gamma$}
\Text(115,45)[]{$U$}
  \end{picture}
\hspace{2cm}
\begin{picture}(80,80)(20,50)
\ArrowLine(90,90)(50,110)
\ArrowLine(50,70)(90,90)
\Line(130,90)(170,110)
\Line(170,70)(130,90)
\Photon(90,90)(130,90){3}{5}
\Photon(68,80)(90,60){3}{6}
\Text(185,110)[]{$f$}
\Text(185,63)[]{$\bar{f}$}
\Text(37,110)[]{$e^{+}$}
\Text(40,70)[]{$e^-$}
\Text(110,100)[]{$\gamma$}
\Text(105,55)[]{$U$}
\end{picture}
\end{center} 
These are examples of $U$strahlung. Because of the smallness of the $U$ boson 
coupling to ordinary particles, such processes are expected to be at least 
$3 \ 10^{-7} \,(m_U/\UUNIT{MeV}{})^2$ times 
lower than the single photon production process ($e^+ e^- \rightarrow f \bar{f} \gamma$). 
These should be difficult to detect for a
light $U$ boson having a mass $\mdm < O(\UUNIT{GeV}{})$.

{}

\end{document}